\documentclass{JASSS}
\usepackage{epsfig}
\usepackage{amsmath,amssymb}
\usepackage{graphicx,xcolor}
\usepackage{hyperref}
\usepackage{soul}

\newcommand{\erefs}[2]{Eqs.~(\ref{e.#1})--(\ref{e.#2})}
\newcommand{\eref}[1]{Eq.~(\ref{e.#1})}
\newcommand{\fref}[1]{Fig.~\ref{f.#1}}

\newcommand{\theor}{\ensuremath{\rm T}}
\newcommand{\parz}[1]{\ensuremath{\left(#1\right)}}
 
\title{Proof of principle for a self-governing prediction and forecasting reward algorithm}

\reviewcopy{false}

\author[1,2]{J.~O.~Gonzalez-Hernandez}
\affil[1]{Dipartimento di Fisica, Universit\`a degli Studi di Torino, Via P. Giuria 1, I-10125, Torino, Italy}
\affil[2]{INFN, Sezione di Torino, Via P. Giuria 1, Torino, I-10125, Italy}

\author[3]{Jonathan Marino}

\author[3]{Ted Rogers}
\affil[3]{Department of Physics, Old Dominion University, Norfolk, VA 23529, USA}

\author[3]{Brandon Velasco}

\email{trogers@odu.edu}

\usepackage{natbib}
	\setcitestyle{authoryear,round,aysep={}}
	
\begin{document}
\maketitle 



\begin{abstract}
We use Monte Carlo techniques to simulate an organized prediction competition between a group of a scientific experts acting under the influence of a ``self-governing'' prediction reward algorithm. Our aim is to illustrate the advantages of a specific type of reward distribution rule that is designed to address some of the limitations of traditional forecast scoring rules. The primary extension of this algorithm as compared with standard forecast scoring is that it incorporates measures of both group consensus and question relevance directly into the reward distribution algorithm. Our model of the prediction competition includes parameters that control both the level of bias from prior beliefs and the influence of the reward incentive. 
The Monte Carlo simulations demonstrate that, within the simplifying assumptions of the the model, experts collectively approach belief in objectively true facts, so long as reward influence is high and the bias stays below a critical threshold. The purpose of this work is to motivate further research into prediction reward algorithms that combine standard forecasting measures with factors like bias and consensus.  
\end{abstract}

\begin{keywords}
Expert judgment, simulations, forecast scoring
\end{keywords}

\parano{}



\section{Introduction}
\label{s.intro}





Prediction competitions have grown rapidly in popularity over the past several decades. Much of their appeal lies with their usefulness in assessing expert judgment and predictive power. An example is the Metaculus project~\citep{metaculus}, a massive online prediction aggregation and assessment engine. Projects such as these incentivize accurate predictions by distributing reputational reward points to participants. In typical prediction competitions of this type, rewards are distributed according to an ordinary proper forecast scoring rule like a Brier score or a logarithmic score. This works well for subjects where predictions tend to have undisputed outcomes and are of obvious relevance to the larger theoretical issues under consideration. However, there are other areas of scientific research that would also benefit greatly from improved methods for systematically collecting and assessing claims of predictive power, but where there has been little interest so far in organized prediction competitions. Often, these are highly technical fields, filled with subtleties that must somehow be addressed before any meaningful assessment of predictive success is possible. Experts in such fields frequently disagree even about the actual outcomes of predictions, or about their relevance to the broader scientific questions under debate. This limits the potential usefulness of straightforward forecast scoring rules, and it makes organized competitive predicting far less attractive to potential participants. A consequence is that true predictive power in these fields remains difficult to quantify, and the situation is especially unclear to external onlookers. 
To assess the outcomes of predictions, or their relevance to broader scientific questions, outsiders are only able to defer to the very experts who made them. The challenges outsiders face in assessing expert judgment~\citep{burgman_trusting_2016} are related to the ``replication crisis'' that has emerged in some areas of science over the past few decades and to other contemporary concerns about scientific methodology~\citep{ritchie_science_2020}.  

Our motivation in this paper is to better understand if there are ways to make competitive prediction aggregation and assessment more feasible in a broader range of subject areas by modifying the reward algorithm.
We explore this by simulating the activities of group of experts acting in accordance with an enhanced reward  algorithm based on a proposal that can be found in \citep{rogers2022self} and which we will review 
below. This enhanced scoring rule combines standard proper forecast scoring with measures of group consensus and prediction relevance, with the aim of fostering  ``adversarial collaboration''~\citep{clark2021adversarial} within the group.  

The simulations that we will present in this paper
are meant to demonstrate, in an idealized but instructive model of $N$ experts who follow the reward algorithm, that when individuals pursue a high net reward the group is lead collectively toward objectively correct conclusions, provided the average bias remains below a certain threshold. We show that this maximum bias threshold can be quite high and still the group tendency toward correct answers remains quite robust. The code for the simulations was created in Wolfram Mathematica~\citep{Mathematica}, and samples of the simulations with documentation are available $\href{https://drive.google.com/file/d/1CE4WNy9XRZHF94WN3XkRW5kGkalddsLl/view}{here}$. Our simulations are intended to establish a basic proof of principle by showing that the reward algorithm performs as it is designed to, at least within the confines of highly idealized model scenarios. Our results also suggest ways that the simulation can be expanded and improved in the future.  Ultimately, we hope that future improvements to the simulations can guide efforts to improve to the reward algorithm itself.  

\section{The Reward Algorithm}
\label{s.rewardalg}
The setup of the reward system discussed in ~\citep{rogers2022self} is that there is a sequence of questions with ``yes'' or ``no'' outcomes. Experts compete to provide the most accurately calibrated probabilistic prediction on each question.  A reward $r_{i,j}$ is given to each expert $i$ for their performance on question $j$.  To motivate the discussion of the details below and organize the discussion, it is useful to begin with the following schematic formula,
\begin{equation}
\label{e.rewars_def}
r_{i,j} \propto \parz{\text{Prediction Accuracy}} \times \parz{\text{Question Significance}} \times \parz{\text{  Consensus on Result}} \, .
\end{equation}
Below, we will step through an explanation of each factor in \eref{rewars_def}.

\subsection{Prediction accuracy}
\label{s.accuracy}

We begin with the first factor in \eref{rewars_def}.
Each expert in the prediction competition is to be rewarded for providing the most accurate (or most calibrated) prediction possible for each question. By this we mean that the probability provided for a ``yes'' outcome should come as close as possible to the true probability, given all available knowledge. With all other factors in \eref{rewars_def} fixed, we might choose the reward to be proportional to a proper forecasting rule. For the algorithm in this paper, we use \emph{surprisal} for ``$\text{Prediction Accuracy}$,'' defined as 
\begin{equation}
\label{e.surprisaldef}
s_{i,j} = \begin{cases} 
	-\ln p_{i,j} \, & \text{if outcome} = \text{yes} \\
	-\ln (1-p_{i,j}) \, & \text{if outcome} = \text{no} \\
	\end{cases} \, ,
\end{equation}
for expert $i$ on question $j$. Surprisal is a proper forecast scoring rule with a number of desirable properties. It is effectively a measure of the quantity of information gained by observing the outcome of a binary question, and it is closely related to the concept of Shannon information content. Note that a low surprisal indicates high predictive power and a large surprisal indicates poor predictive power. A surprisal of zero indicates perfect predictive power.

A limitation of \eref{surprisaldef} with respect to a \emph{self-governing} algorithm is that the experts in real life debates frequently disagree about what the outcomes of predictions actually are (see the discussion in the introduction). In the absence of external referees, it is necessary to have some proxy for the yes or no ``outcome'' on the right side of \eref{surprisaldef}. We will defer to the expert wisdom of the crowd for this and define what we call the ``resolution'' of expert $i$ on question $j$ as 
\begin{itemize}
\item $v_{i,j} = +1$  if expert $i$ believes or asserts that the outcome was ``yes'' 
\item $v_{i,j} = -1$ if expert $i$ believes or asserts that the outcome was  ``no'' 
\item $v_{i,j} = 0$  if expert $i$ supplies no answer \, .
\end{itemize}
The mean of all experts' resolutions regarding question $j$ is then
\begin{equation}
\label{e.meanV}
V_j = \frac{1}{N_j} \sum_i^{N_j} v_{i,j} \, ,
\end{equation}
where $N_j$ is the total number of experts who supplied predictions on question $j$. Finally, let us call the \emph{effective} outcome $q_j$ for question $j$
\begin{equation}
q_j = \begin{cases}
	\text{yes (+1)}  \,& \text{if} \;\;  V_j > 0 \\
	\text{no (-1)} \, & \text{if} \;\;  V_j < 0 \\
	0 \, & \text{if} \;\;  V_j = 0 \, 
\end{cases} \, .  \label{e.outcome_def}
\end{equation}
Now we may write a less ambiguous version of 
\eref{surprisaldef}, appropriate for our purposes, 
\begin{equation}
\label{e.surprisaldefp}
s_{i,j} = \begin{cases} 
	-\ln p_{i,j} \, & \text{if} = q_j = +1 \\
	-\ln (1-p_{i,j}) \, & \text{if} = q_j = -1  \\
	\end{cases} \, .
\end{equation}
Of course, the reliability of this measure of predictive success now depends on the reliability of the group consensus. 

Since some question outcomes will be much more difficult to predict than others, the reward given to an expert for their accuracy should depend on their surprisal \emph{relative} to that of their peers, rather than on their absolute surprisal. To quantify this, we calculate the mean, the mean-squared, and the standard deviation of all surprisals for all experts on question $j$, 
\begin{align}
\langle s_j \rangle &= \frac{1}{N_j} \sum_i^{N_j} s_{i,j} \, , \label{e.meansur} \\
\label{e.stdsur}
\langle s_j^2 \rangle &= \frac{1}{N_j} \sum_i^{N_j} s_{i,j}^2 \, \\
\Delta s_j &= \sqrt{\langle s_j^2 \rangle-\langle s_j \rangle^2} \, .
\end{align}
An expert's reward should be large if their surprisal is far below what can be considered a large surprisal on question $j$. The exact size of suprisal that we consider ``large'' here is somewhat arbitrary, but typically it will be some number $c$ of standard deviations above the mean surprisal $\langle s_j \rangle$ of the group on question $j$. Therefore, we define a big surprise on question $j$ to be
\begin{equation}
s^{\text{Big}}_j = \langle s_j \rangle + c \Delta s_j \, . \label{e.bigsur}
\end{equation}
We will fix the exact numerical value for $c$ later.

To summarize, we will make the reward for player $i$ on question $j$ proportional to the distance of their surprisal below the big surprisal $s^{\text{Big}}_j$,
\begin{equation}
\label{e.accuracyreward}
r_{i,j} \propto s^{\text{Big}}_j - s_{i,j} \, .
\end{equation}
The right side of \eref{accuracyreward} is what will use for the first factor in \eref{rewars_def}.

\subsection{Question significance}
\label{s.relevance}

The purpose of the reward algorithm is to incentivize adversarial collaboration in the prediction making process. However, if all experts are equally surprised by an outcome, that is if $\Delta_j \approx 0$ on question $j$, then the critical adversarial component is missing. The reward system should incentivize question-prediction sequences that tend to  demonstrate one set of ideas' predictive advantages over anothers'. Thus, to account for the second factor in \eref{rewars_def} we also make the reward for expert $i$ on question $j$ proportional to the overall $\Delta s_j$ for that question, 
\begin{equation}
\label{e.sigreward}
r_{i,j} \propto \Delta s_j \, .
\end{equation}
An expert can expect to obtain a large reward only if $\Delta s_j$ is reasonably large. The group of experts cannot use trivial or irrelevant predictions to inflate their overall reward count.

\subsection{Consensus regarding results}
\label{s.consensus}

It is frequently the case that real life experts do not agree on 
whether a particular prediction was confirmed or refuted by observations or measurements.  Sometimes this is because the original questions were vaguely formed. In other cases, there is disagreement over the details of how measurements were made or how data should be interpreted. This limits the reliability of the surprisal by itself, as it is calculated in \eref{surprisaldefp}, as a measure of predictive accuracy. Robust collaborative efforts are necessary to avoid such scenarios and ensure that a reasonably broad consensus is established for each question and prediction cycle. Therefore, the amount of reward distributed should be weighted less if there is only weak consensus.   

The absolute value of the mean resolution in \eref{meanV} is a quantitative measure of the consensus. If all experts agree that the outcome of prediction $j$ was ``yes,'' then $V_j = +1$, while if all agree that it was ``no,'' then $V_j = -1$. If half of experts believe the outcome was ``yes'' and the other half believe it was ``no,'' then $V_j = 0$, and there was no consensus. Thus, to account for the last factor in \eref{rewars_def}, we will make the reward proportional to $|V_j|$,
\begin{equation}
\label{e.consreward}
r_{i,j} \propto | V_j | \, .
\end{equation}
We will refer to $| V_j |$ as the ``consensus'' for question $j$. A consensus of 1 means that all experts agree the result was either ``yes'' or ``no.'' A consensus of 0 means the experts are exactly evenly split. In the latter case, there is no way to determine a surprisal, and no reward points are given out.

\subsection{Summary: the reward formula}
\label{s.formula}

To summarize the above, the reward for expert $i$ on question $j$ is to be simultaneously proportional to $s_j^\text{Big} - s_{i,j}$ (\eref{accuracyreward}), $\Delta s_j$ (\eref{sigreward}), and $|V_j|$ (\eref{consreward}). Substituting these into \eref{rewars_def} gives our final version of the reward formula, 
\begin{equation}
r_{i,j} = \parz{s_j^\text{Big} - s_{i,j}} \Delta s_j |V_j| \, , \label{e.reward_formula}
\end{equation}
up to an overall factor that fixes one unit of reward. This is the quantitative version of \eref{rewars_def}. Note that a standard logarithmic forecast scoring rule would include only the first factor. 

In order for any individual expert to accumulate a large reward, the group must regularly reach consensus through collaboration. Otherwise, $|V_j|$ will tend to be small. There must also be an adversarial element to each question-prediction cycle.  Otherwise, $\Delta s_j$ will be small and again the ability for any expert to accumulate a significant reward will be limited. Finally, each expert must supply an earnest and carefully considered prediction for each question or risk having a trend of small $s_j^\text{Big} - s_{i,j}$. Thus, \eref{reward_formula} rewards the full constellation of behaviors we set out to incentivize with \eref{rewars_def}. 

Summing \eref{reward_formula} over all players gives the total amount of reward distributed for question $j$, 
\begin{equation}
r_{j,\text{total}} = \sum_{i = 1}^{N_j} r_{i,j} = c N_j \Delta s_j^2 |V_j| \, . \label{e.reward_total}
\end{equation}
This equation provides an interpretation for $c$. It fixes the average reward per expert on question $j$. If $c$ is very large and positive, then all experts will tend to receive at least some positive reward, even for relatively inaccurate predictions. If $c = 0$, then the total reward handed out is zero, and it becomes a zero sum prediction competition. In that situation, every reward point earned by one expert is accompanied by a negative reward (a reward penalty) for another.  We will fix $c = 1$ for now. This ensures that only a consistent pattern of extremely poor predictions can lead to a net negative reward for an expert. 

After $n$ predictions, the \emph{total} reward accumulated by expert $i$ will be represented with a boldface $\mathbf{r}_i$,
\begin{equation}
\mathbf{r}_i \equiv \sum_{j=1}^n r_{i,j} \, .
\end{equation}
Of course, each expert will try to maximize their own $\mathbf{r}_i$.

\section{A basic demonstration}
\label{s.tests}

Let $\theor$ stand for a specific theoretical belief or hypothesis that can either be true or false. ($\theor$ stands for ``theory.'') For the purposes of illustration, we will assume that $\theor$ is objectively true and that belief in it therefore increases an expert's ability to predict the outcomes of general predictable questions. 


\subsection{Belief and prediction accuracy}
\label{s.beliefaccuracy}

The basic setup of the simulation is that there are $N$ experts competing on each question, and there are $n$ total questions with ``yes'' or ``no'' outcomes.  In our simulations, we will generate the outcomes $\omega_j$ for the questions randomly, with $\omega_j = 1$ corresponding to ``yes'' and $\omega_j = -1$ corresponding to ``no.''

Each player $i$ has a list of attributes, the most basic of which is their level of belief in $\theor$. If they have a strong belief in $\theor$, then their probabilistic forecasts $p_{i,j}$ for $\omega_j$ will tend, on average, to be close to 1 when $\omega_j = +1$ and close to zero when $\omega_j = -1$. If they actively \emph{disbelieve} $\theor$, then the situation is reversed; their forecasts will tend to be close to 0 when $\omega_j = +1$ and close to 1 when $\omega_j = -1$. For simplicity, we will quantify degree of belief $d_i$ discretely with 
\begin{equation}
\label{e.degreeofbelief}
d_i = \begin{cases} 
    +4 \, & \text{strongest belief}  \\
    +3 \, & \text{strong belief} \\
	+2 \, & \text{moderate belief} \\
	+1 \, & \text{weak belief}  \\
    0 \, & \text{undecided}  \\
    -1 \, & \text{weak disbelief}  \\
    -2 \, & \text{moderate disbelief}  \\
    -3 \, & \text{strong disbelief}  \\
    -4 \, & \text{strongest disbelief}  \\
	\end{cases} \, .
\end{equation}
To generate a forecast for expert $i$ on question $j$, we will 
first generate a random real number $\nu$ in the range $\left( 0,1\right)$. Then, 
\begin{equation}
\label{e.modelforecast}
p_{i,j} = \begin{cases} p(\nu,d_i) & \text{if} \;\; \omega_j = +1 \, (\text{true}) \\ 
             1 - p(\nu,d_i) & \text{if} \;\; \omega_j = -1 \, (\text{false})
             \end{cases} \, ,
\end{equation}
where $p(\nu,d_i)$ is a function that takes values between 0 and 1. We will use simple power laws for $p(\nu,d_i)$,
\begin{equation}
\label{e.forecastprofile}
p(\nu,d_i) = 
\begin{cases} 
1 - \nu^{21} & d_i = 4 \, \\
1 - \nu^{5.3} & d_i = 3 \, \\
1 - \nu^{2.7} & d_i = 2 \, \\
1 - \nu^{1.6} & d_i = 1 \, \\
1 - \nu & d_i = 0 \, \\
\nu^{1.6} & d_i = -1 \, \\
\nu^{2.7} & d_i = -2 \, \\
\nu^{5.3} & d_i = -3 \, \\
\nu^{21} & d_i = -4 \, \\
\end{cases} \, .
\end{equation}
The power of $\nu$ for each degree of belief in  \eref{forecastprofile} is chosen so that the average probabilities associated with each $d_i$ are separated by even increments of approximately $11.4\%$. An undecided expert ($d=0$) will assert an average probability of  $p(\nu,d=0)=50\%$, $d = 1$ will have an average $p(\nu,d=1)=61.4\%$, and so on up to approximately $95\%$ for $d = 4$. Degrees of disbelief ($d < 0$) are the mirror image of the degrees of positive belief, so that the average $p(\nu,-d)$ is the average $1-p(\nu,d)$. The questions that  make up the prediction competition are intended to have at least some degree of uncertainty, so we will prohibit experts from asserting probabilities above or below some threshold. In our simulations of the next section, this threshold will be $1.0\%$, so that $p_{ij}\in [0.01,0.99]$.

\begin{figure*}[htp]
	\begin{center}
	\includegraphics[width=12cm]{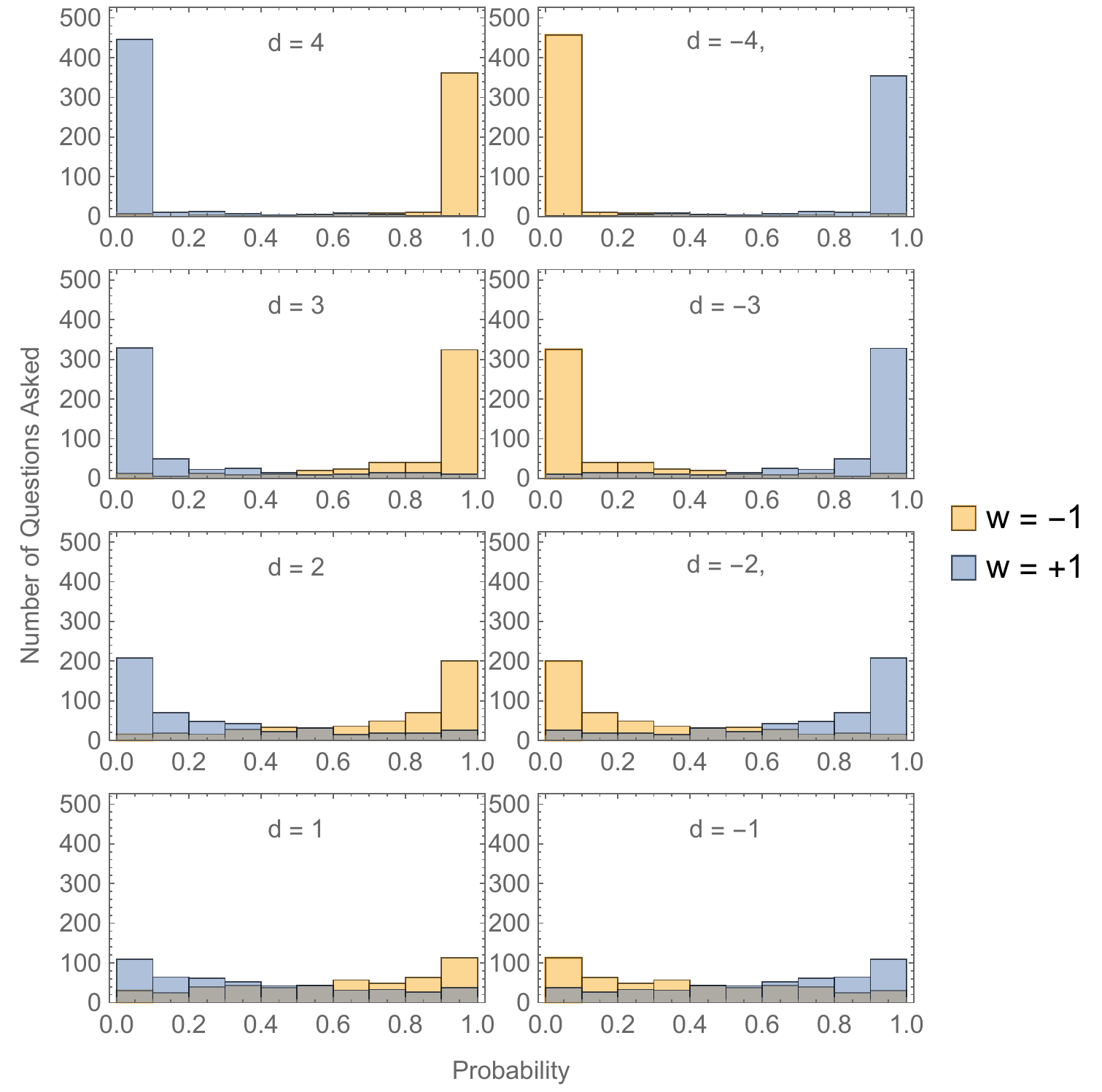}
\end{center}
	\caption{Typical histograms of forecasts for each nonzero value of $d$ in \eref{modelforecast}.}
	\label{fig:histograms}
\end{figure*}

Histograms illustrating typical forecasts are shown Fig.~\ref{fig:histograms} for different degrees of belief. 
Note that they are symmetric under a change in $\theor$ from true to false. In the types of scenarios we are considering, each question might involve a complicated mixture of different beliefs aside from $\theor$. A significant belief in $\theor$ pushes an expert toward highly calibrated predictions on average, but it does not guarantee an accurate prediction for each question. This means experts will be somewhat hesitant to update their beliefs, even after receiving low rewards, particularly if they are each aware of the presence of bias throughout the group.   

\subsection{Updating beliefs}
\label{s.updating}

After each question-prediction-outcome cycle, there is some chance that each expert will update their belief in $\theor$. To mimic typical human expert behavior, we will simulate updating by separating the process into the following two steps:

\subsubsection{Random walk updating}

The baseline scenario is that experts are highly unlikely to update their beliefs. In the absence of any external motivating factor, an expert will only increase or decrease their degree of belief randomly and with a small probability. 
We will call this probability the ``mutation rate'' and we will label it by $\mu$. The value of $\mu$ is fixed to be between $0$ and $1$. Then, the simulation first generates a random number $\nu$ and if $\nu < \mu$ it assigns $d_{i,j+1} = d_{i,j} + 1$ (unless $d = 4$). It then generates another random number $\nu$ and if $\nu < \mu$ it assigns $d_{i,j+1} = d_{i,j} - 1$ (unless $d = -4$). Therefore, if there were no other updating, each expert's belief would be a random walk. In most of our simulations, we will use $\mu = 0.01$. Thus, after each prediction an expert has a $2\%$ chance of shifting their belief up or down one unit. Given a small mutation rate, the beliefs of the experts will slowly drift away from an initial consensus until they are totally randomized. 

\subsubsection{Reward motivated updating}
\label{s.rwu}

In our simulations, our aim is to observe what happens when we tie belief updates to the reward formula in \eref{reward_formula}. 
The basic assumption behind the simulation is that an expert with a low reward will tend to realign their belief to more closely match that of the experts with higher rewards. To express this in symbols, let $i = M_j$ correspond to the expert with the highest (maximum) total accumulated reward at question $j$. Then, $\mathbf{r}_{M_j,j}$ is the \emph{maximum} reward held by any expert by question $j$, and $d_{M_j,j}$ is the degree of belief held by the player with the maximum reward.
Other experts will update their beliefs if their reward falls some distance below what might be considered a ``large'' reward. 
Depending on the personality traits of the specific expert, a large reward might mean anything above the mean reward, or it could be something close to $\mathbf{r}_{M_j,j}$. We will compromise between these two ways that experts might compare themselves to their peers and define a ``large'' reward to be the weighted average of $\mathbf{r}_{M_j,j}$ and $\langle \mathbf{r}_j \rangle$,
\begin{equation}
\label{e.largereward}
\mathbf{r}_j^\text{Large} =  x \langle \mathbf{r}_j \rangle + y \mathbf{r}_{M_j,j} \, ,
\end{equation}
with $x + y = 1$. 
An individual expert's ``reward deficit'' $\delta \mathbf{r}_{i,j}$ will be how far below $\mathbf{r}_j^\text{Large}$ their accumulated reward at question $j$ falls, 
\begin{equation}
\delta \mathbf{r}_{i,j} = \mathbf{r}_j^\text{Large} - \mathbf{r}_{i,j} \, .
\end{equation}
Each expert will have a high probability of updating their belief 
if their reward deficit is a critical fraction of $\mathbf{r}_j^\text{Large}$. The farther their reward falls below the large reward, the less likely they are to retain their prior belief. 
We implement this with the formula,
\begin{align}
d_{i,j+1} 
= \begin{cases} d_{i,j} & \nu \leq e^{-a_j \frac{\delta \mathbf{r}_{i,j}}{\mathbf{r}_j^\text{Large}}} \\
d_{i,j} + \text{sign}(d_{M_j,j}- d_{i,j}) & \;\text{otherwise}
\end{cases} \, , \label{e.affinity}
\end{align}
where $\nu$ is again a random real number between 0 and 1.  Here, $a_j$ is a real positive parameter that we will call the ``affinity'' of the experts. If $a_j$ is very large, the experts with relatively low reward relative to their peers will very quickly shift their beliefs to match that of the experts with the largest rewards. 

Since experts are only likely to update beliefs in situations where reasonably large rewards are available, we will not choose a fixed number for $a_j$, but instead write
\begin{equation}
a_j = a_0 \frac{r_{j,\text{total}}}{r_{j,\text{total}} + r_0}
\label{e.affinity0}
\end{equation}
The value of $r_0$ is a threshold reward that we will fix later. 
If the total award distributed on question $j$ is large relative to $r_0$, then the affinity is $a_j \approx a_0$. If the total reward is close to zero, then $a_j \approx 0$.

\subsection{Modeling consensus and bias} 
\label{s.reachingconsensus}

If the objective outcome $\omega_j$ of a prediction turns out not to match an expert's expectations, they may be inclined to reject the objective outcome and instead choose a $v_{i,j}$ that is more in line with their prior beliefs. This bias effect also needs to be modeled in our simulations. 
To do so, we assume that if expert $i$'s surprisal is much larger than the mean on question $j$, then they will reject the objective outcome and choose instead $v_j = - \omega_j$.

To express this in a formula, define a surprisal $\hat{s}_{i,j}$ that is calculated exactly as in \eref{surprisaldefp}, but now using $\omega_j$ rather than $q_j$. (We can call this the ``objective surprisal'' since it is based on the objective underlying reality rather than the expert consensus.) Then, we determine the validation number of expert $i$ on question $j$ by using
\begin{equation}
v_{i,j} = \begin{cases} \omega_j & \text{if}\;\; \nu \leq e^{ -b \left( \frac{\hat{s}_{i,j}}{\langle \hat{s}_j \rangle + b_0}\right)}  \\
-\omega_j & \text{otherwise}
\end{cases} \, . \label{e.biaseq}
\end{equation}
$\nu$ is again a randomly generated real number between 0 and 1. Modeling the effects of bias involves two real, positive parameters, $b$ and $b_0$. 

If, on one hand, the average objective surprisal is large, $\langle \hat{s}_j \rangle \gg b_0$, then all or most of the experts have made an inaccurate prediction, and so they likely hold the wrong belief. In this case, it is only $b$ that matters in \eref{biaseq}. In this case there is a probability of $\approx e^{-b}$ that an expert with a typical objective surprisal will agree with the objective outcome. Thus, $b$ represents a collective, group bias. 

On the other hand, if the group of experts is mostly correct in their predictions, so that $\langle \hat{s}_j \rangle \ll b_0$, then  $\langle \hat{s}_j \rangle$ becomes irrelevant in 
\eref{biaseq}. Only the handful of experts with poor predictions will be likely to reject the objective outcome, and their probability for accepting the objective outcome is 
$\approx e^{-b/b_0}$. Thus, $b/b_0$ is a measure of individual biases against a correct group consensus.

For moderate $\langle \hat{s}_j \rangle$, \eref{biaseq} interpolates between the two scenarios above. 

We will simply call the parameter $b$ the ``bias'' of the experts. Since we are mainly interested in the effect of collective bias for now rather than individual biases, we will set 
\begin{equation}
\label{e.biasthreshold}
b_0 = 0.7 \, .
\end{equation}
(A surprisal of $s \approx 0.7$ is the result of a $p \approx 0.5$ forecast.) We will study the effect of changing $b_0$ below. 
If $b$ is large, then expert $i$ will only match their $v_{i,j}$ to $\omega_j$ when their objective surprisal is small relative to $\hat{s}_{i,j} \ll \langle \hat{s}_j \rangle + b_0$. If $b \approx 0$, then the experts will almost always agree with the correct, objective outcome. If $b \approx \infty$, the experts will always disagree with the objective outcome whenever $\hat{s}_j > \langle \hat{s}_j \rangle$. 

\subsection{Exit criterion}

In real-world scenarios, groups of experts will usually cease to question an underlying theory $\theor$ once beliefs have stabilized and the potential for large rewards has vanished. To simulate this, we will halt the questions once 
\begin{equation}
\label{e.stopcondition}
r_{j,\text{total}} < r_{j,\text{threshold}}
\end{equation}
for $n_\text{stable}$ consecutive questions. 
In cases where stability is not reached, we exit the simulation after $n_\text{max}$ questions.
Later, we will estimate the value of $r_{j,\text{threshold}}$ based on the behavior of 20 experts in a prediction competition. 

\subsection{Parameter selection}
\label{s.parameters}

In our sample simulations, we choose parameters to represent reasonable expectations for the behavior of an actual group of experts competing to make accurate predictions. For statistical measures like average reward or average belief to be meaningful, the number of experts should be large but still realistic for an actual community of experts in a highly specialized subject. It also must be kept small enough for simulation times to be manageable.  
We will take $N_j = 20$. (If a typical small expert research group publishes articles with around 5 authors, then this would corresponds to a competition between about 4 research groups.) If enough rewards points are distributed so that there is at least an approximately $10\%$ probability per expert that a belief will be modified, then there is roughly an $88\%$ chance that with each question a belief will be updated. If there is only a random walk, with a $0.2\%$ chance for each expert to update on each question, then there is roughly a $33\%$ chance that at least one expert's belief will be updated on each question. Thus, over the course of several questions, we expect the distinction between $2\%$ per expert (random walk) and $\approx 10\%$ per expert to become evident. If there were a $\lesssim 1/3$ chance of an $r_{j,\text{total}} < r_{j,\text{threshold}}$ entirely by random chance, then there a $\lesssim 0.01\%$ of having $r_{j,\text{total}} < r_{j,\text{threshold}}$ on three questions consecutively. Having $n_\text{stable} \approx 4$ questions in a row where there is a very small reward is, therefore, a reliable indication that the group experts have converged upon a upon a stable degree of belief. 

We also estimate that $a_0 \approx 0.1-0.2$ is a reasonable affinity for describing a typical expert. To see why, consider an expert with $\delta \mathbf{r}_{i,j} \approx \mathbf{r}_j^\text{Large}$ on question $j$, and assume that $r_{j,\text{total}} \gg r_0$. Then from~\eref{affinity} the probability that the expert shifts their belief $d$ by one increment is approximately $10\%-40\%$. Repeated over the course of five questions, this scenario yields a probability between $41\%$ and $97\%$ for the expert to increment their belief $d$. 

For the affinity related quantities in \eref{largereward}, we make the choices
\begin{align}
\label{e.choicesxyr0}
    x={}&y=\frac{1}{2},\qquad r_0=50.
\end{align}
We have estimated the value of $r_0$ by considering the situation where, out of 20 experts, half of them have the strongest believe in theory $\theor$ ($d=4$), and the other half have the strongest disbelief ($d=-4$). Keeping fixed the degree of belief, we simulate an experiment with a large number of questions and calculate the total rewards $r_{j,\text{total}}$ at every step. The average over all questions is $\langle r_{\text{total}}\rangle\approx 100$, which is an estimate for very large rewards given in a single question. Since this estimate applies to a rather extreme scenario, we take half of this value, $r_0=\langle r_{\text{total}}\rangle/2$. We will estimate the effect of adjust $x$ and $y$ in the robustness section below. 

For the exit criterion, we set
\begin{align}
\label{e.choicesexit}
r_{j,\text{threshold}} ={}&4.04,\,\,
n_{\text{stable}}=4,\,\,
n_{\text{max}}=1000 \, .\nonumber
\end{align}
As with $r_0$, the value of  $r_{j,\text{threshold}}$ was estimated by simulating an experiment where all experts agree on $d=4$. In this case, we obtain $\langle r_{\text{total}}\rangle\approx2.02$, an estimate for very small rewards given in a single question. Since we do not necessarily require that \emph{all} experts agree on $d=4$ before exiting, we allow the threshold reward to be somewhat larger, so we set $r_{\text{threshold}}=2\langle r_{\text{total}}\rangle$. 

For estimating reasonable ranges for the bias parameter $b$, 
let us recall that when an expert's surprisal is typical, $\hat{s}_{i,j} \approx \langle s_j \rangle + b_0$, the probability that they will agree with the objective outcome is $\approx e^{-b}$. Thus, an expert with $b \lesssim 0.7$ is relatively unbiased while one with $b \gtrsim 0.7$ is somewhat biased. 

\section{Numerical examples}
\label{s.examples}

\begin{figure*}
\centering \includegraphics[scale=0.32,trim=0  0 0 0]{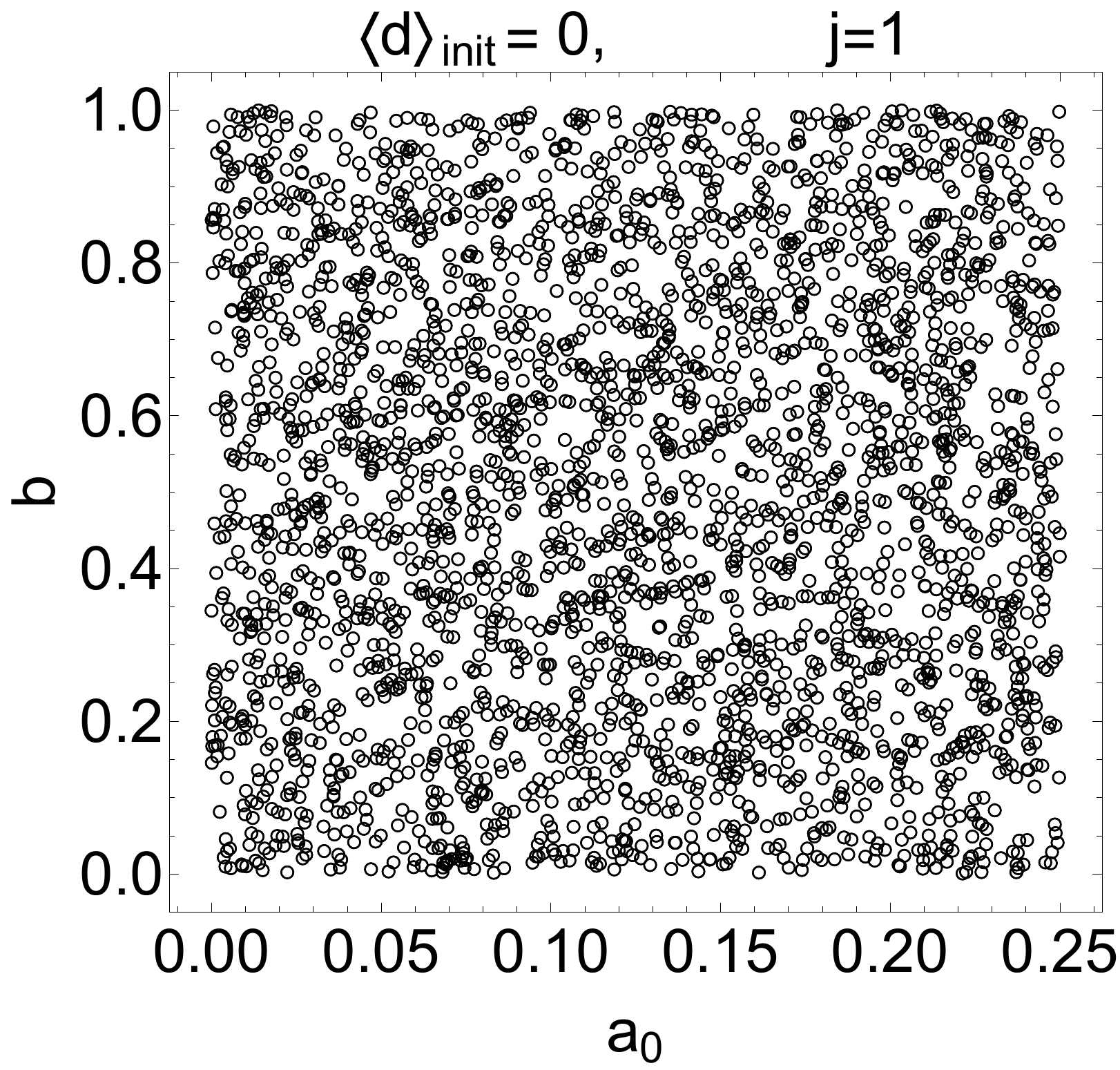}
    \includegraphics[scale=0.32,trim=0  0 0 0]{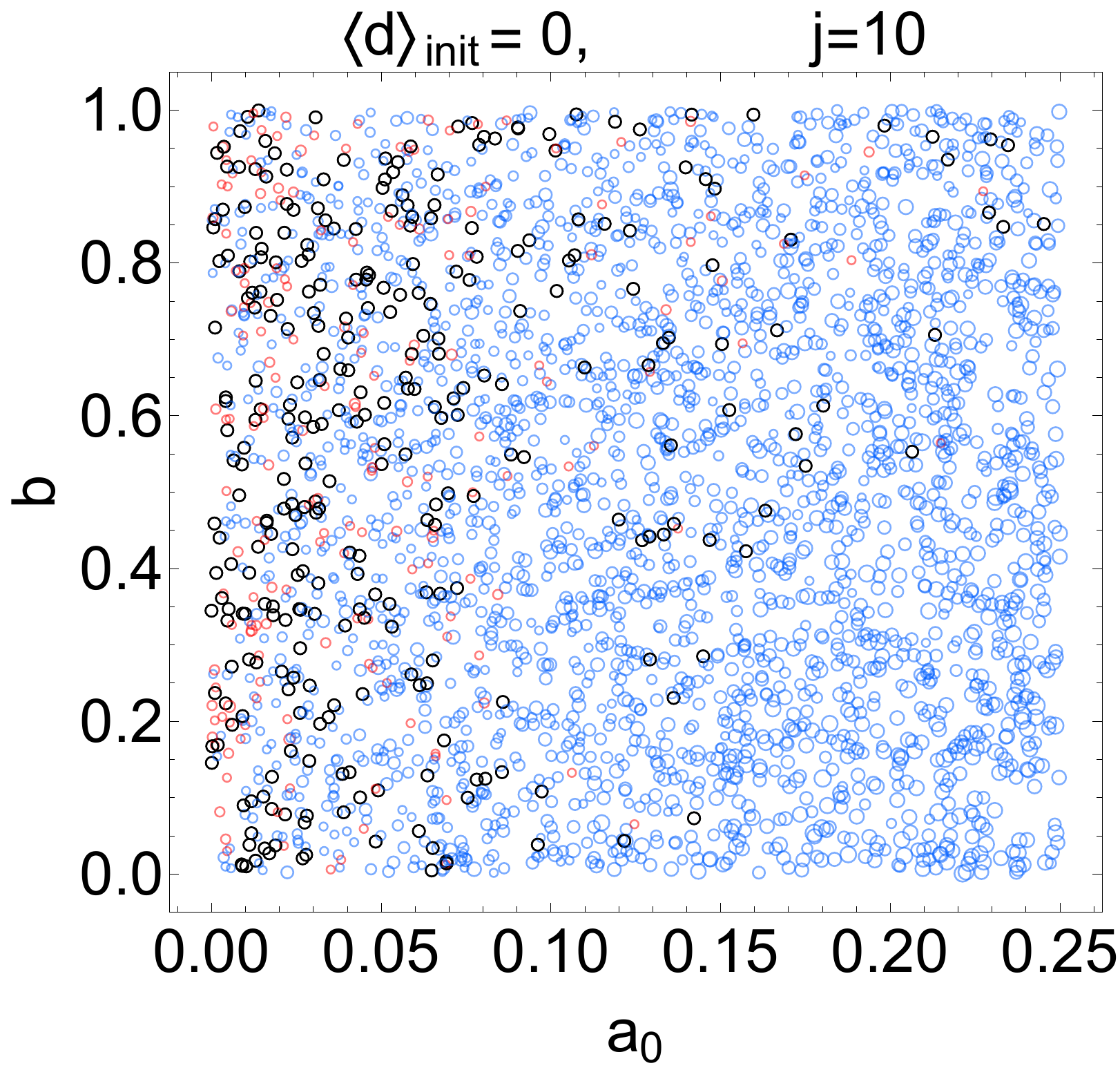}\\
    \includegraphics[scale=0.32,trim=0  0 0 0]{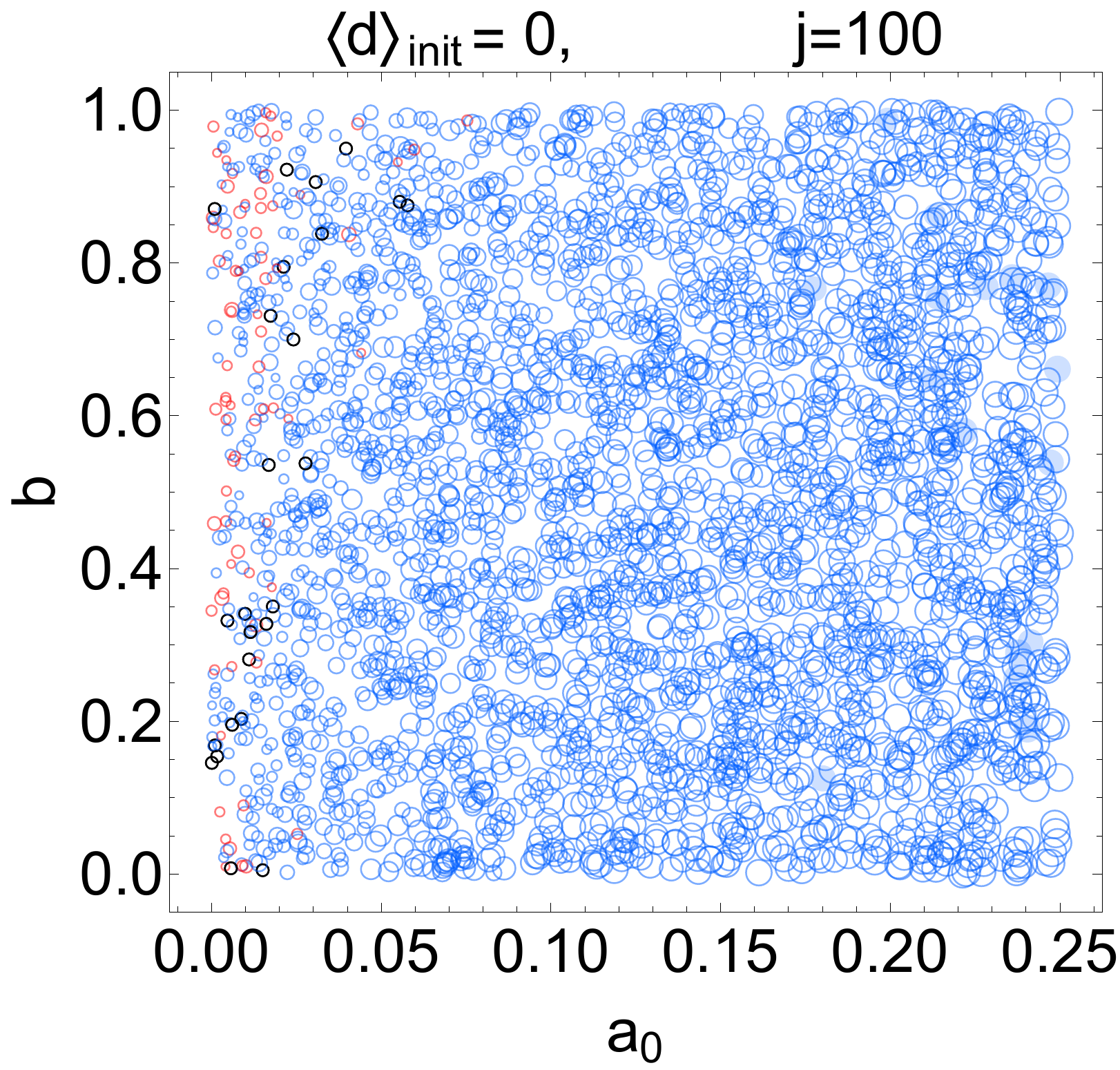}
    \includegraphics[scale=0.32,trim=0  0 0 0]{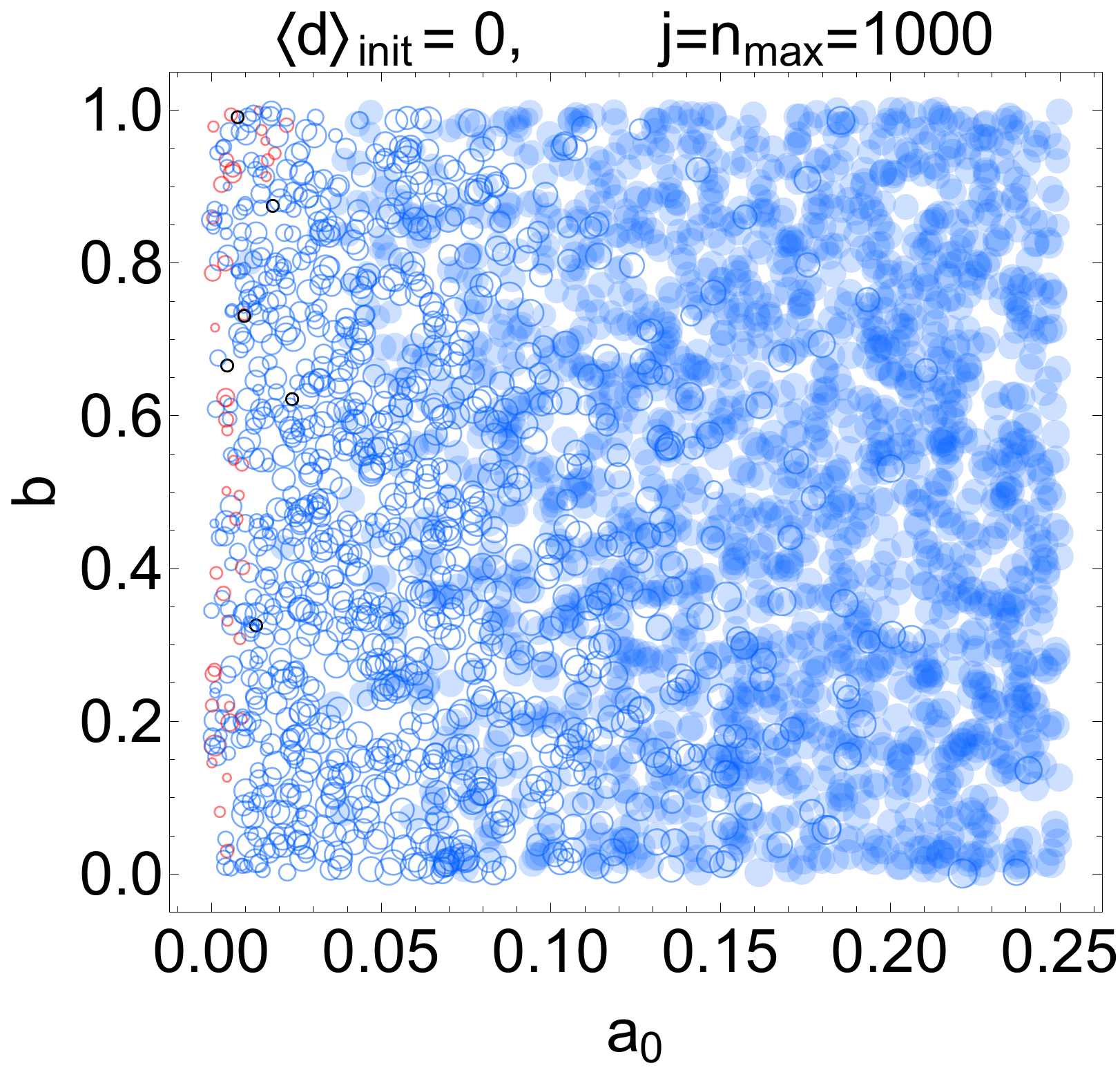}
    \caption{Each circle in the panels above represents a single simulation of $j$ questions with a group of 20 experts. The position of the circle in the plane indicates one pair of values (randomly chosen) for $b$ and $a_0$. Red circles are simulations that result in $\langle d \rangle < 0$ after $j$ questions, blue circles are simulations that result in $\langle d \rangle > 0$ after $j$ questions, and black circles are where $\langle d \rangle = 0$ after $j$ simulations. In all of these simulations, half of experts begin with a strong belief in $\theor$ ($d = +4$) and half begin with a strong disbelief in $\theor$ ($d = -4$).}
    \label{f.phase_caseI} 
\end{figure*}
\begin{figure*}
\centering \includegraphics[scale=0.32,trim=0  0 0 0]{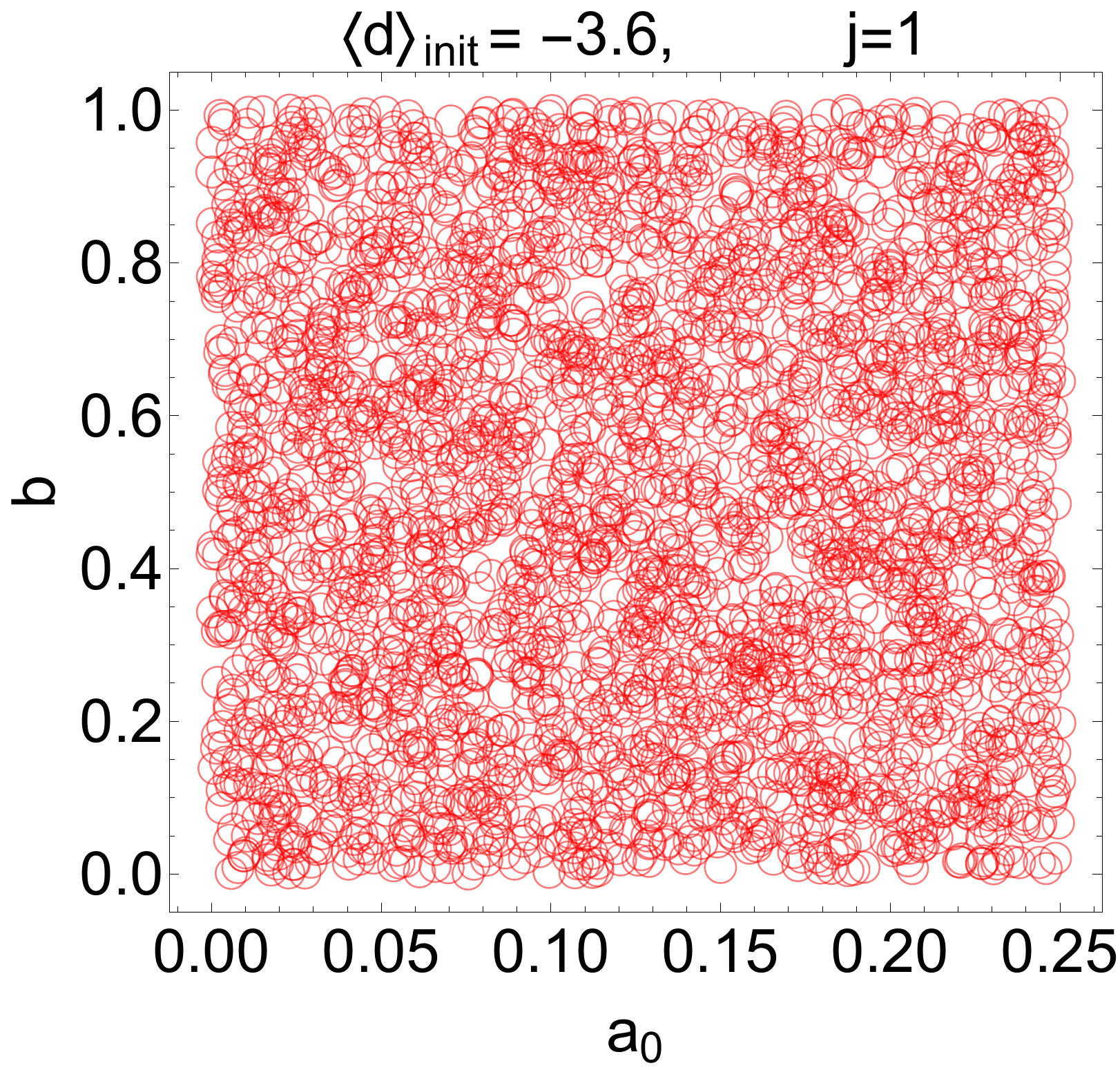}
    \includegraphics[scale=0.32,trim=0  0 0 0]{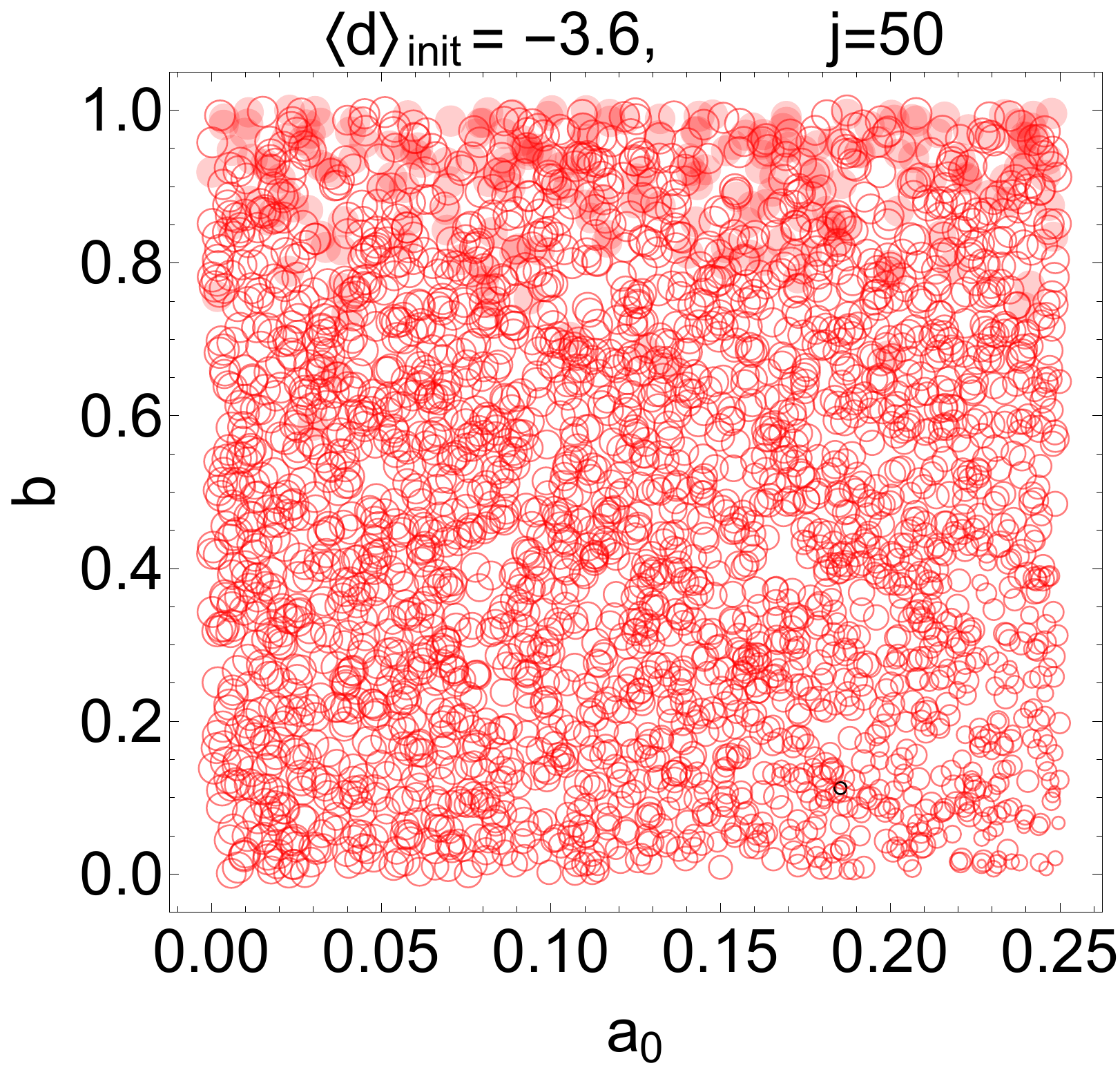}\\
    \includegraphics[scale=0.32,trim=0  0 0 0]{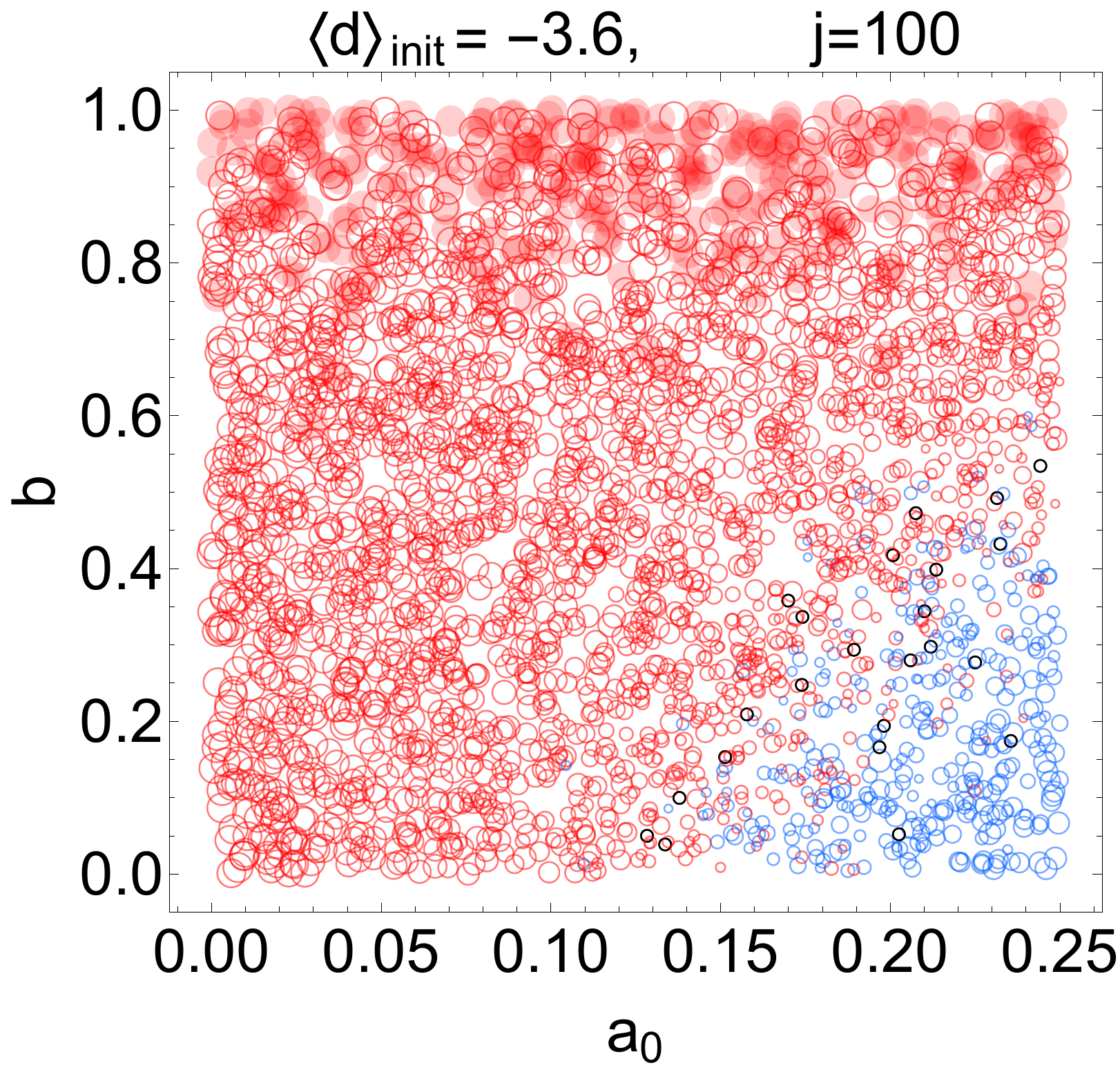}
    \includegraphics[scale=0.32,trim=0  0 0 0]{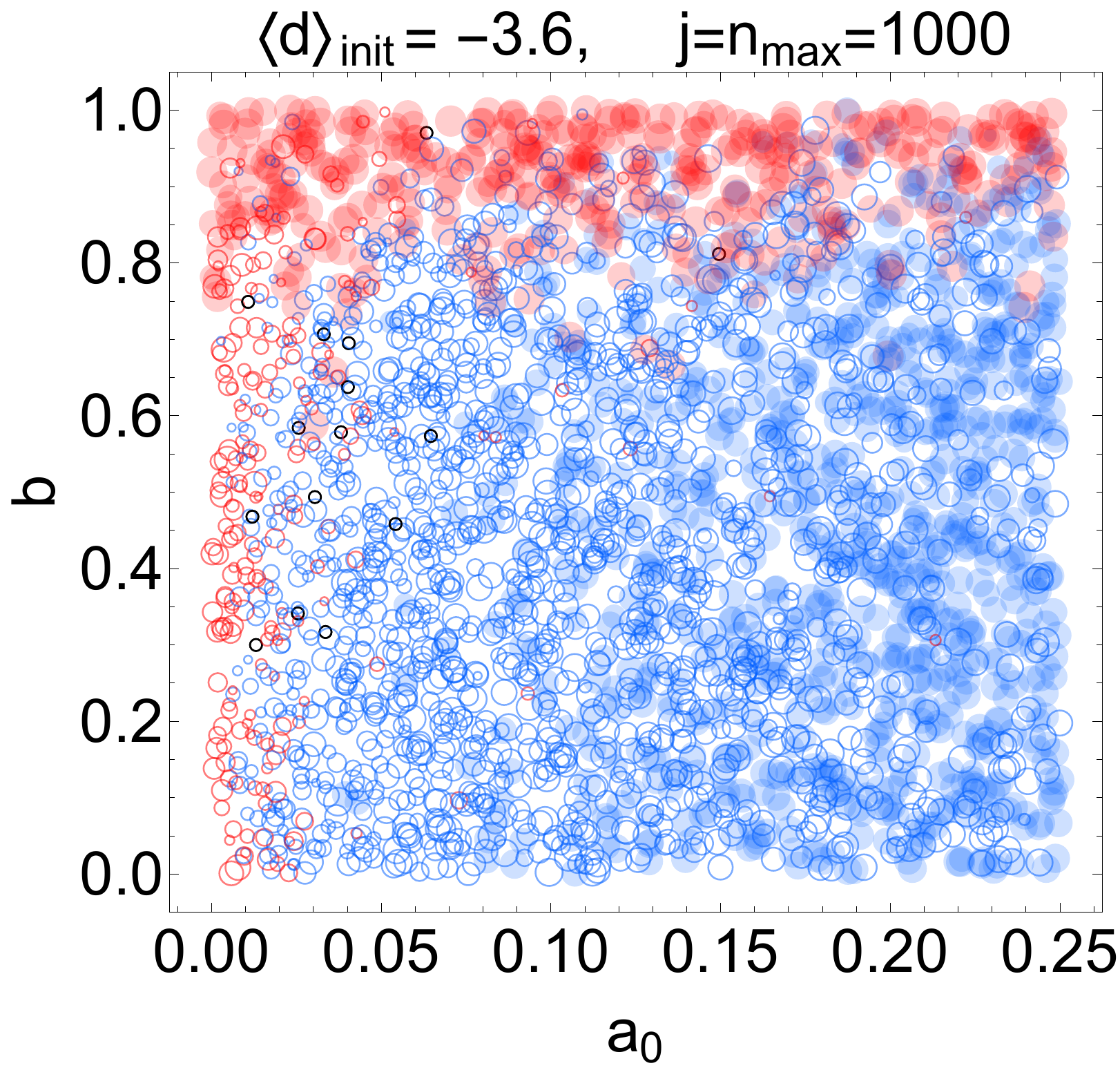}
    \caption{Same as \fref{phase_caseI}, but now 19 of the 20 experts start with strong \emph{dis}belief ($d = -4$) in $\theor$. Only one expert starts with a strong belief ($d = +4$).} 
    \label{f.phase_caseII} 
\end{figure*}

Figures~\ref{f.phase_caseI}-\ref{f.phase_caseII} show the outcomes of simulations with 20 experts, with each panel corresponding to a different average initial $d$.
The vertical axes in each panel indicates the bias parameter $b$ and the horizontal axes show the basic reward affinity parameter $a_0$. Each red or blue circle represents the outcome of a single simulation, and its position in the graph is given by the values of $b$ and $a_0$ used in that simulation.  A blue circle means that the final average $d$ at the end of the simulation was positive (a correct belief in the truth of $\theor$), while a red circle means the average belief is negative (disbelief in $\theor$). A black circle means that the final average degree of belief was $d=0$. If the stop condition was reached in a simulation, i.e. if the inequality in~(\ref{e.stopcondition}) was satisfied for $n_{\text{stable}}=4$ consecutive questions, we indicate this with a solid, filled circle.
A large blue circle indicates a strong average final belief ($d$ close to 4) while a small blue circle indicates weak average belief ($d$ closer to 0). An analogous interpretation applies to the size of the red circles, with red indicating degrees of \emph{dis}belief. 
The different panels in each figure show snapshots at intermediate steps of the 
simulations. For instance, \fref{phase_caseI} shows the state of belief at questions $j=1,10,100$ and $j=n_{\text{max}}=1000$. 

Figure~\ref{f.phase_caseI} shows the outcomes of $3000$ numerical simulations in a scenario where half of the $20$ experts start with $d = -4$ (strong disbelief) and half start with $d = +4$ (strong belief). We use this to establish a baseline, since at a minimum the reward distribution algorithm should cause the group of experts to migrate toward the correct answer if they begin with evenly split beliefs in the correctness of $\theor$. Figure~\ref{f.phase_caseI} confirms this basic trend. With a large affinity $a_0 \gtrsim 0.15$ and a small bias $b \lesssim 0.45$, the group quickly migrates toward belief in $\theor$ after only a handful of questions. After $j=1000$ questions, the group always settles on belief in $\theor$, even when there is a large bias. 

Figure~\ref{f.phase_caseII} is a more interesting case. There, all but one of the 20 experts begin with a strong disbelief ($d = -4$) in $\theor$. A single contrarian expert has a $d = +4$ belief. Naturally, many predictions are required to shift the belief of the majority into the $d > 0$ region, but if $a_0 \gtrsim 0.15$ and $b \lesssim 0.4$, a shift is essentially guaranteed to begin at least around the $100$ question mark. Even after 50 questions, the sizes of the red circles have begun to shrink in the lower right-hand corner of the plot, indicating that belief in $\theor$ has started to become more evenly split. 
As the number of predictions approaches infinity, the group is guaranteed to converge on belief in $\theor$ so long is $a_0$ and $b$ lie within the largely blue rectangular region of the $j = 1000$ plot. Notice that, if $b \gtrsim 0.85$, the group is unlikely to ever abandon its disbelief in $\theor$ regardless of how many predictions are made or how many questions are asked. 

Here it is worth recalling what $a_0$ and $b$ quantify at the level of individual experts. The reward affinity $a_0$ quantifies the experts' willingness to update their level of belief in $\theor$ based on their predictive accuracy relative to their peers if they assume that the reward is an accurate measure of predictive power. Since that last assumption is nontrivial, and since it depends on the ability for the group to reach consensus and agree on the outcomes of predictions, then the value of $a_0$ also quantifies the experts' overall trust in the reward system. The value of $b$ quantifies the experts' tendency to disagree about the objective outcomes of each question/prediction cycle. Figure.~\ref{f.phase_caseII} shows that when the group has a large enough $a_0$, the system can tolerate a relatively large bias $b$ and still migrate toward the objectively correct conclusion ($\theor$ is true). Conversely, as long as $b$ is very small, the group will migrate to the correct conclusion even with only a modest $a_0$. The reliability only breaks down completely when there is a high bias (large $b$) and/or very little trust in the system (low $a_0$). 

If \emph{all} the experts begin with strong disbelief, then one should expect to require very many questions (and very low bias) before there is a shift in belief. 
Conversely, if more of the experts begin with $d = 4$ ($\theor = \text{true}$), then one should expect the transition from $\langle d \rangle < 0$ to $ \langle d \rangle \geq 0$ to take place much faster. These trends are confirmed by \fref{phase_casesII023}, where three additional scenarios are considered: $i)$~all experts start at $d=-4$ (top row), $ii)$~2 experts start at $d=4$ and the rest at $d=-4$ (central row), $iii)$~3 experts start at $d=4$ and the rest at $d=-4$ (bottom row). Scanning from top to bottom, the transition from disbelief to belief (at large $a_0$ and small $b$) happens at earlier $j$. If $3/20$ of the experts begin with a strong belief in $\theor$ (bottom row)
, and $b \lesssim 0.25$, $a_0 \gtrsim 0.2$, the transition begins already by $j = 50$.  Importantly, the top row
confirms that the transition from disbelief to belief eventually occurs even if \emph{all} the experts start with strong disbelief. All that is required is that $b \lesssim 0.6$. 

\begin{figure*}
\centering \includegraphics[scale=0.32,trim=0  0 0 0]{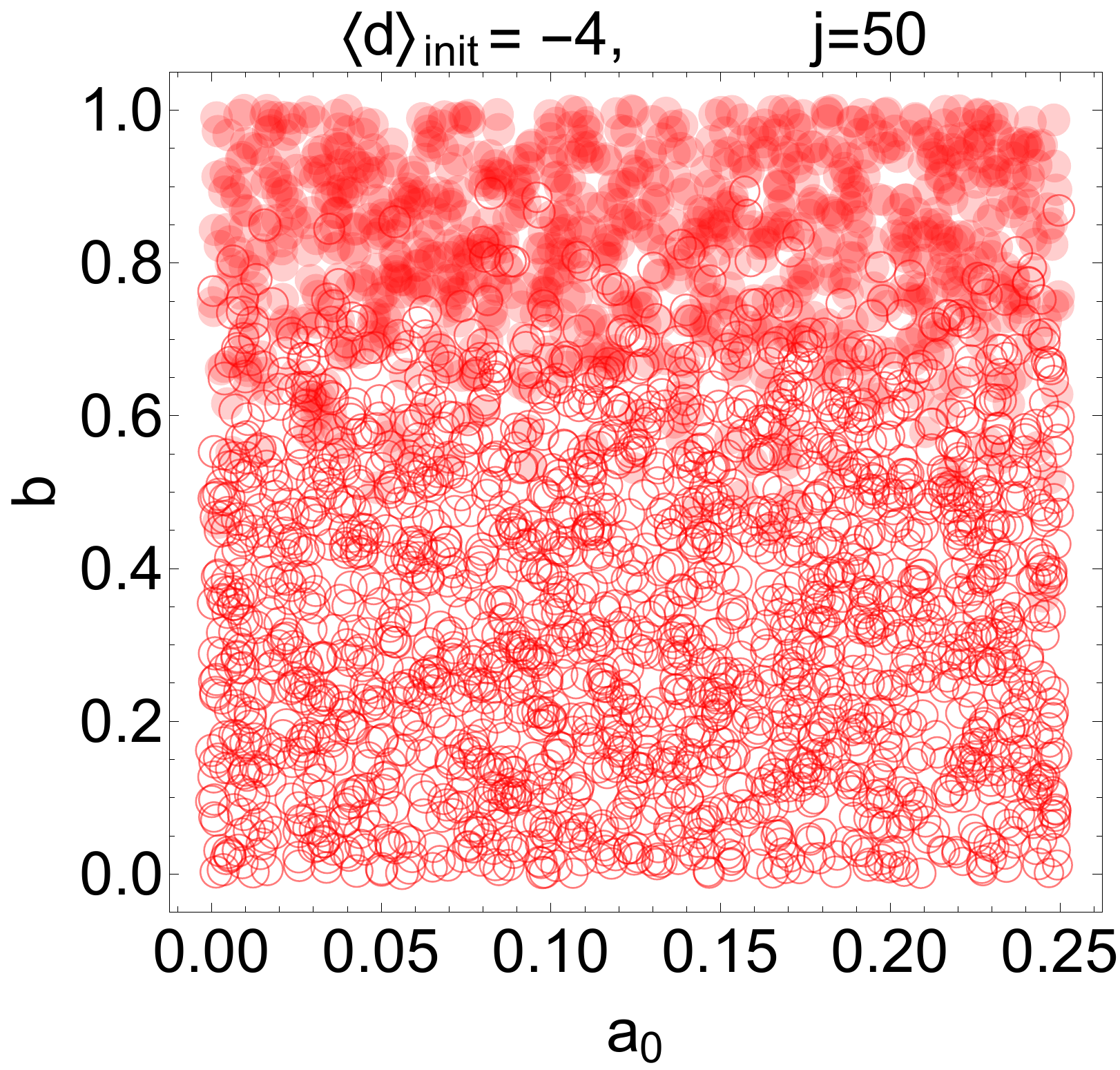}
\includegraphics[scale=0.32,trim=0  0 0 0]{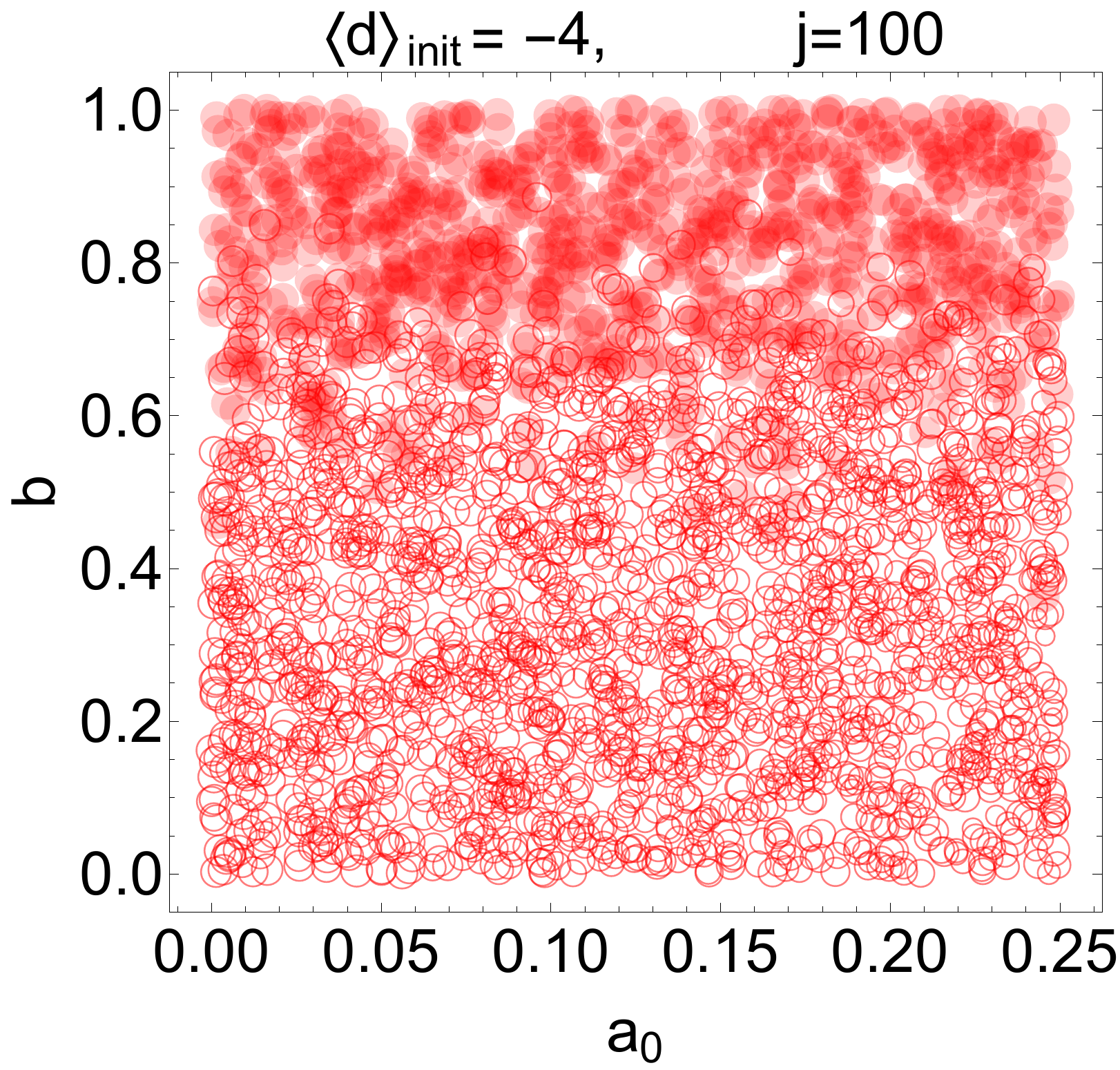}
\includegraphics[scale=0.32,trim=0  0 0 0]{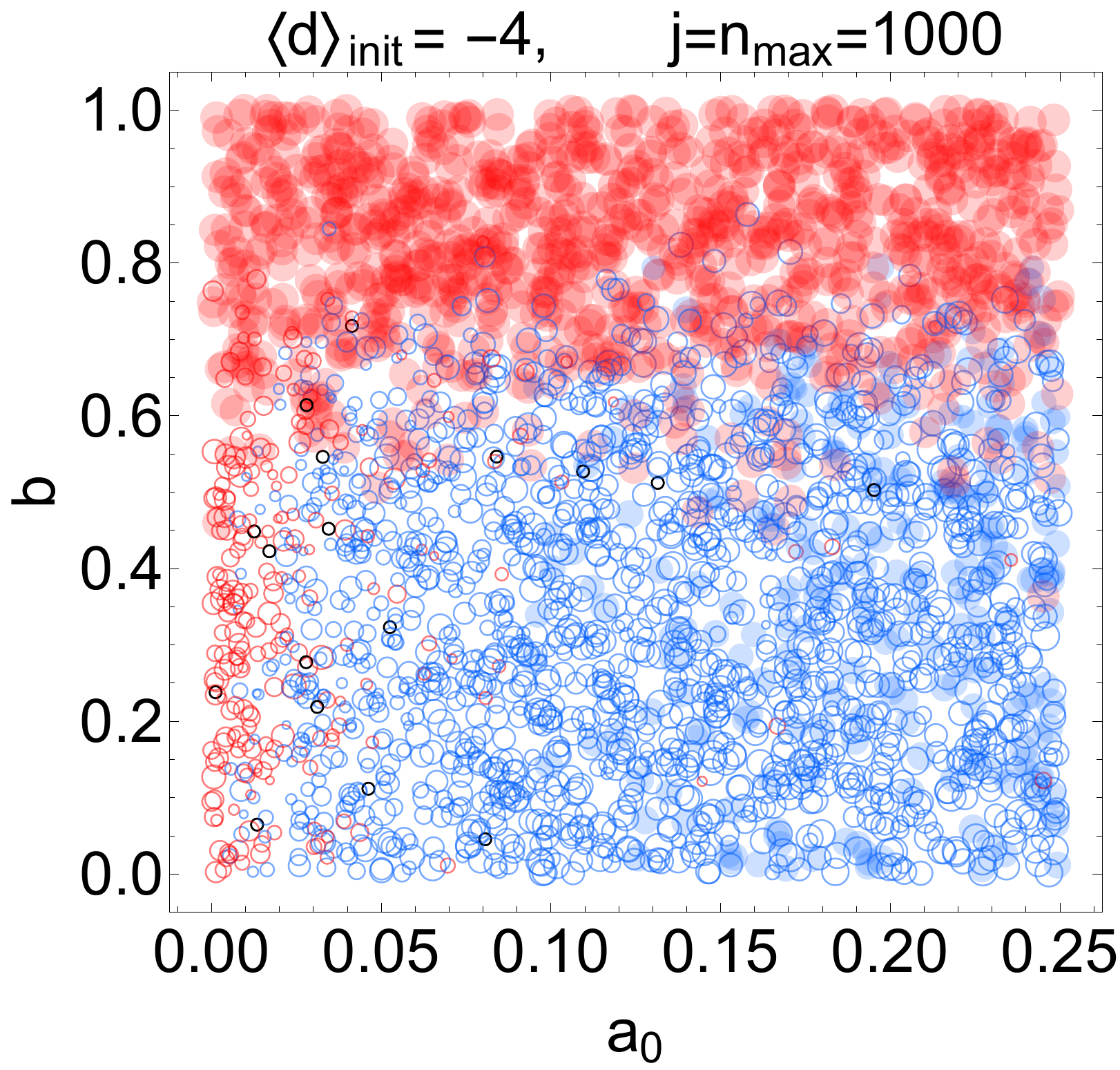}
\\
\includegraphics[scale=0.32,trim=0  0 0 0]{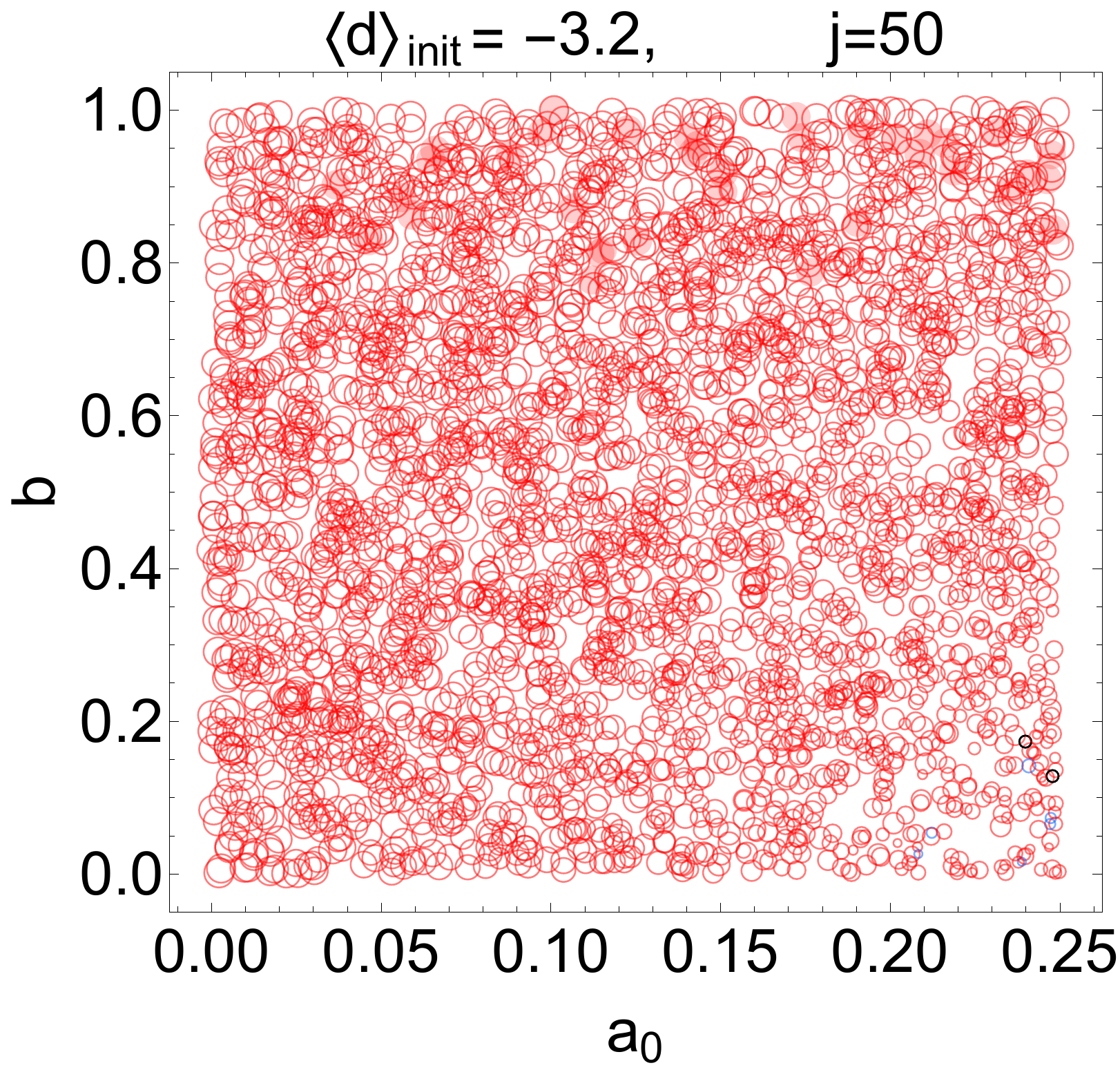}
\includegraphics[scale=0.32,trim=0  0 0 0]{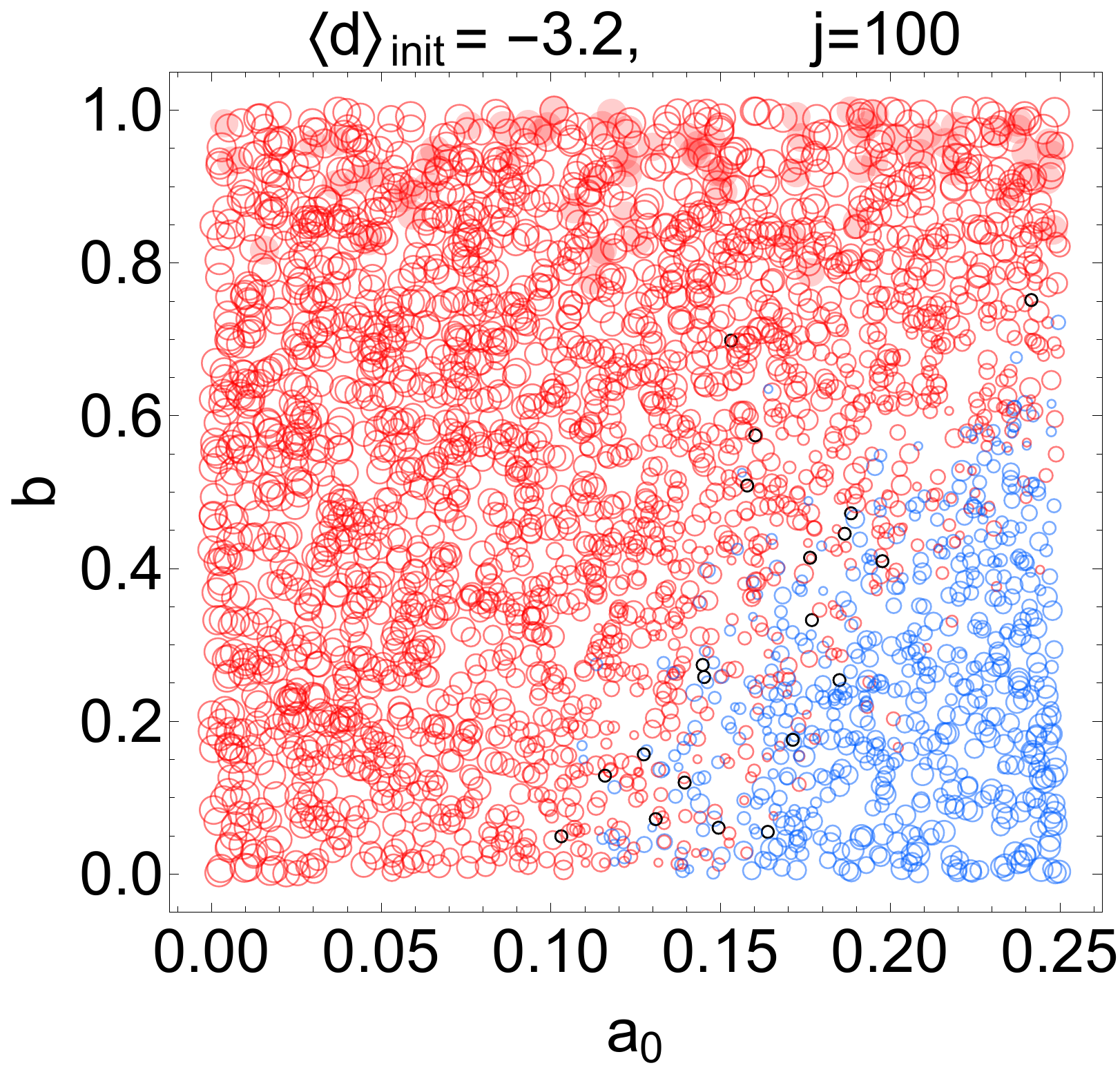}
\includegraphics[scale=0.32,trim=0  0 0 0]{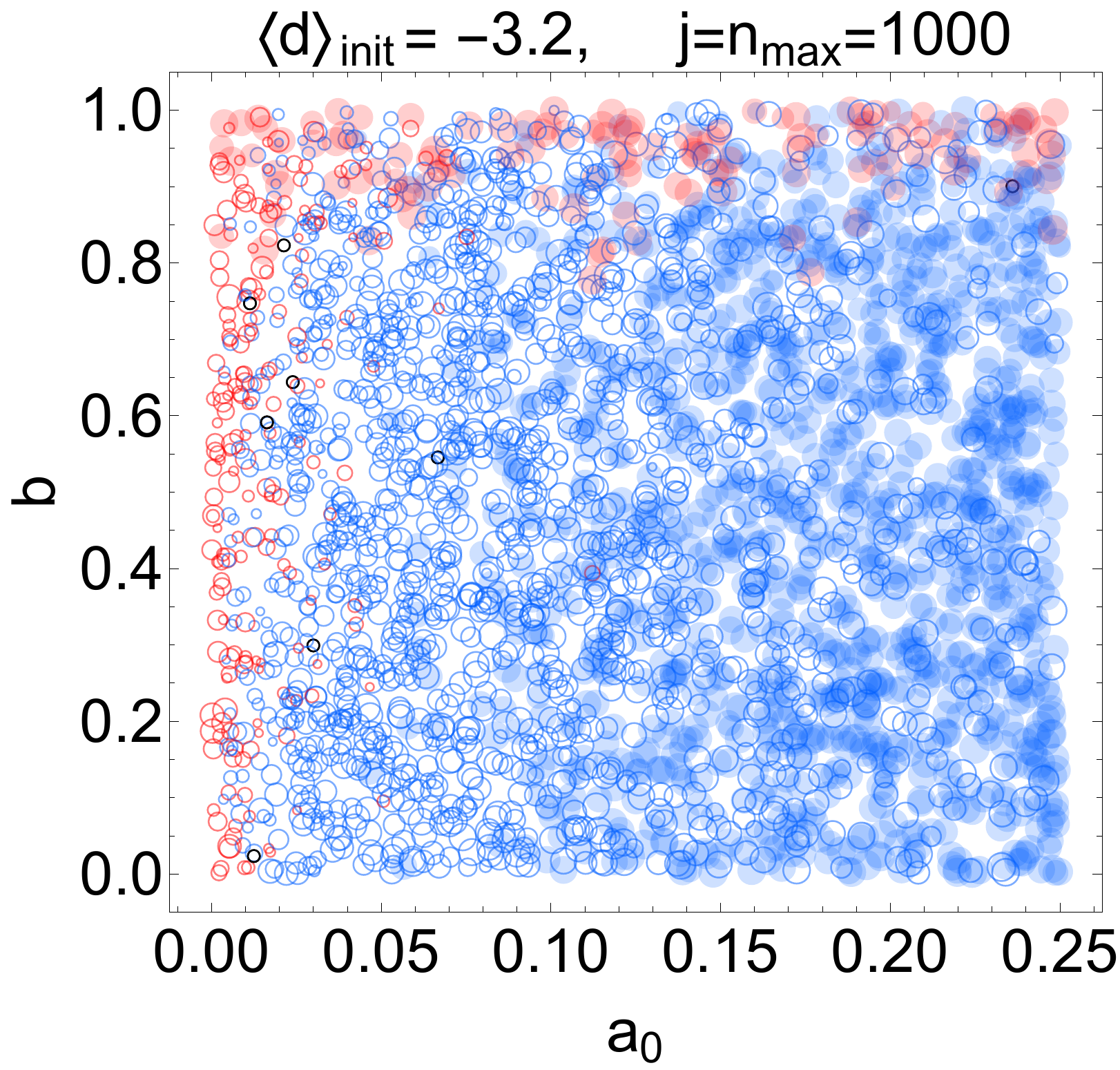}
\\
\includegraphics[scale=0.32,trim=0  0 0 0]{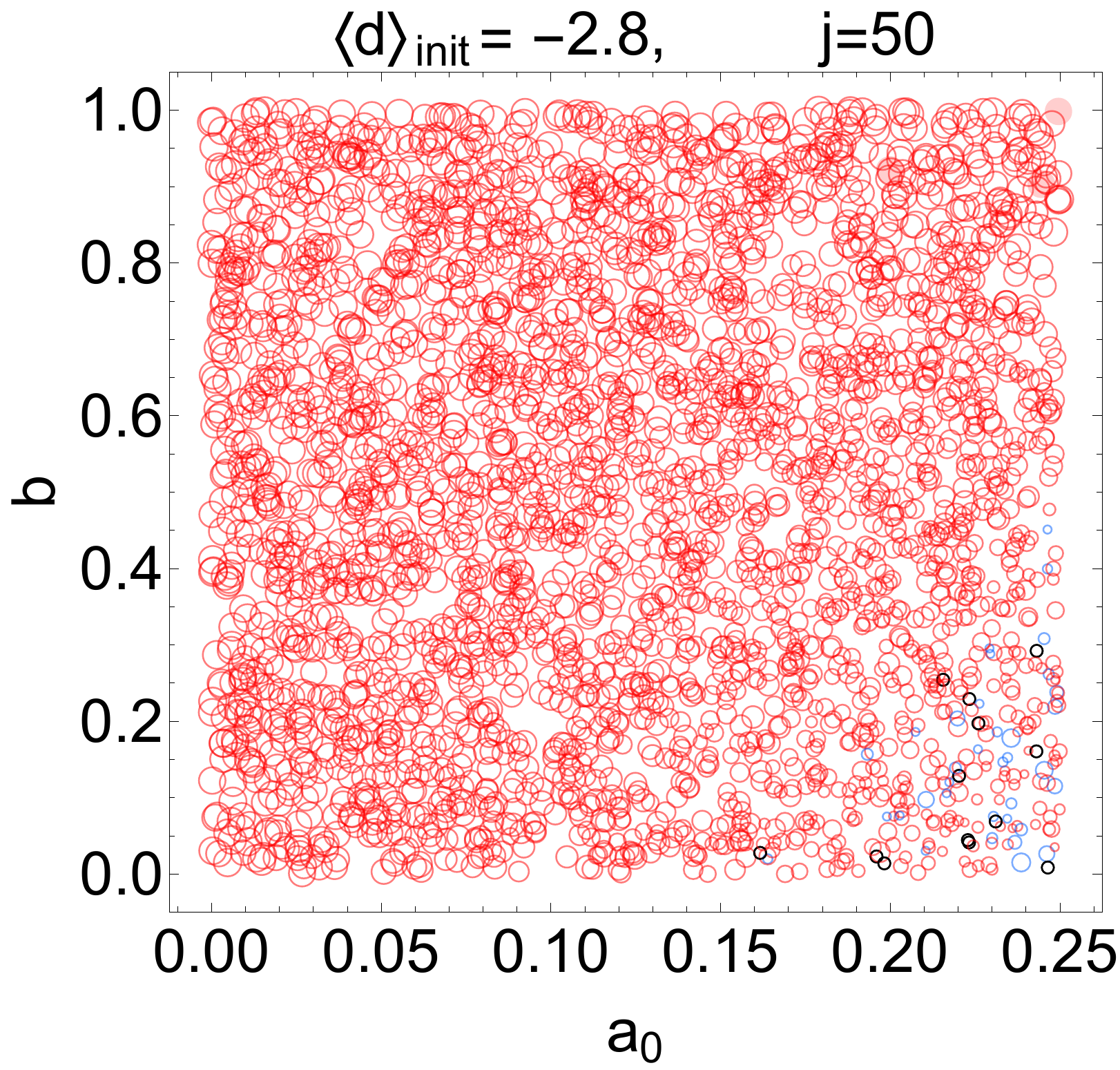}
\includegraphics[scale=0.32,trim=0  0 0 0]{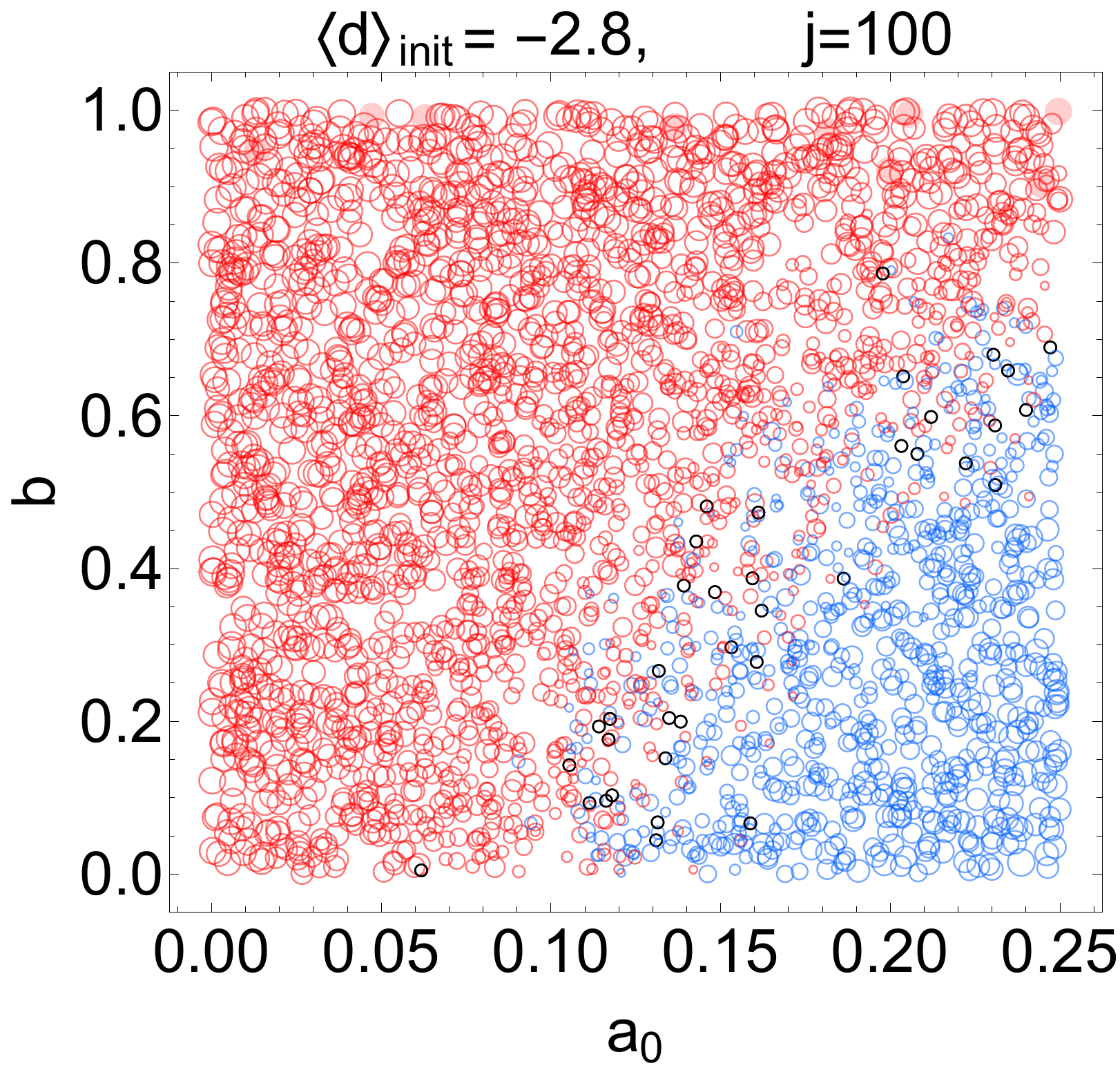}
\includegraphics[scale=0.32,trim=0  0 0 0]{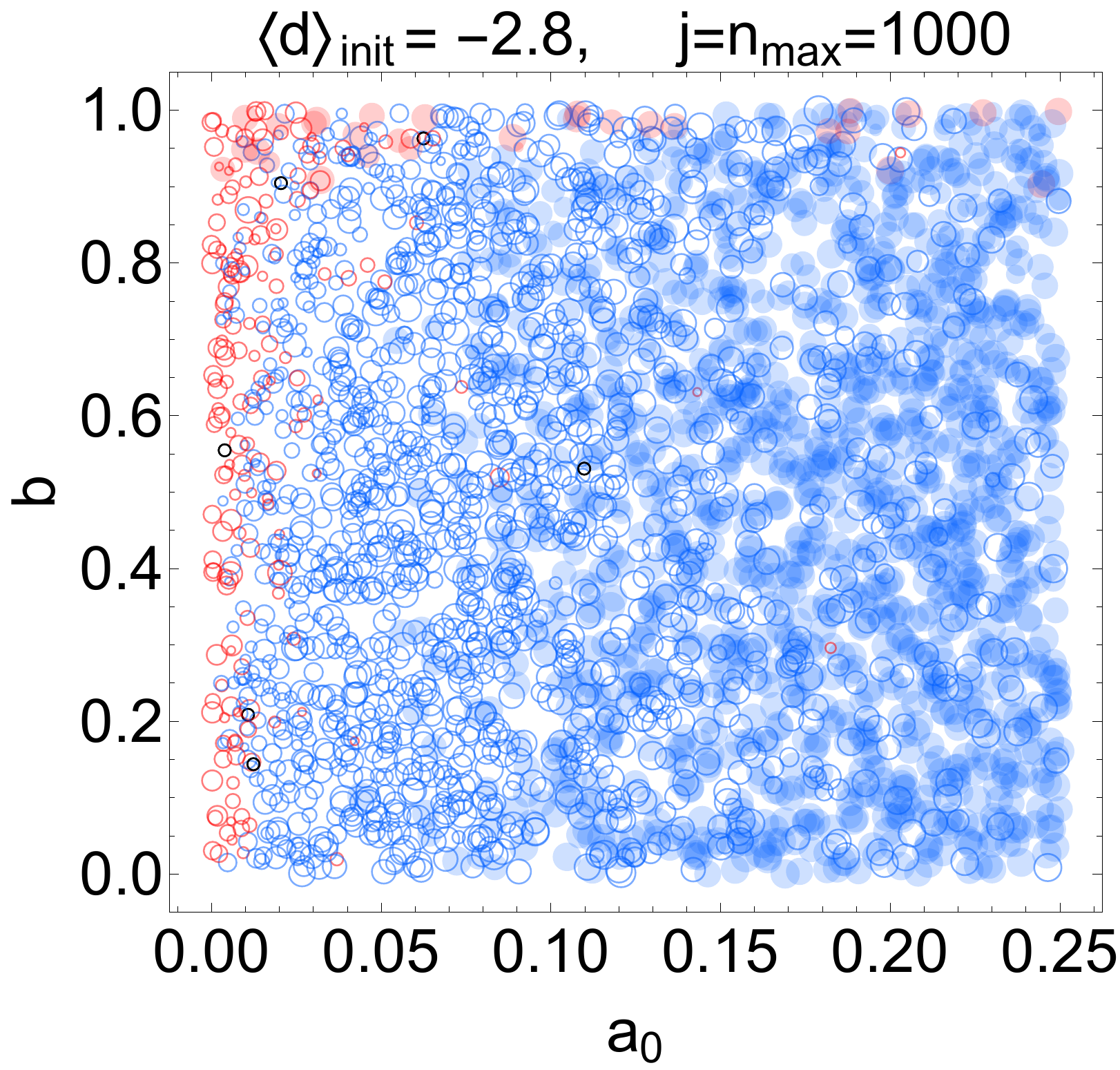}
    \caption{Additional permutations of the initial belief state: $i)$~all experts start with $d=-4$ (top row), $ii)$~2 experts start with $d=4$ and the rest with $d=-4$ (central row), $iii)$~3 experts start at $d=4$ and the rest at $d=-4$ (bottom row).
    }
    \label{f.phase_casesII023} 
\end{figure*}
\begin{figure*}
\centering 
\includegraphics[scale=0.32,trim=0  0 0 0]{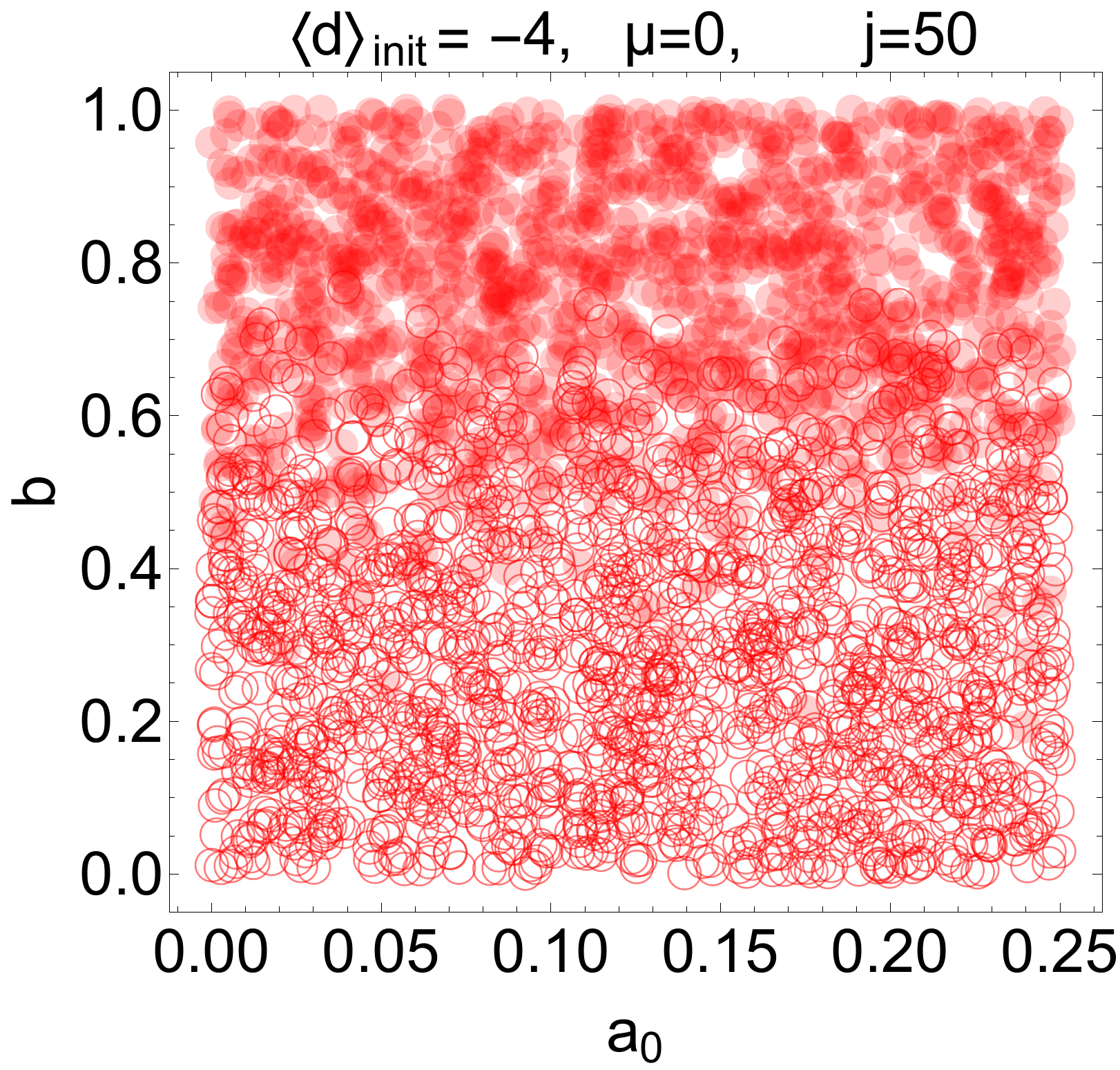}
\includegraphics[scale=0.32,trim=0  0 0 0]{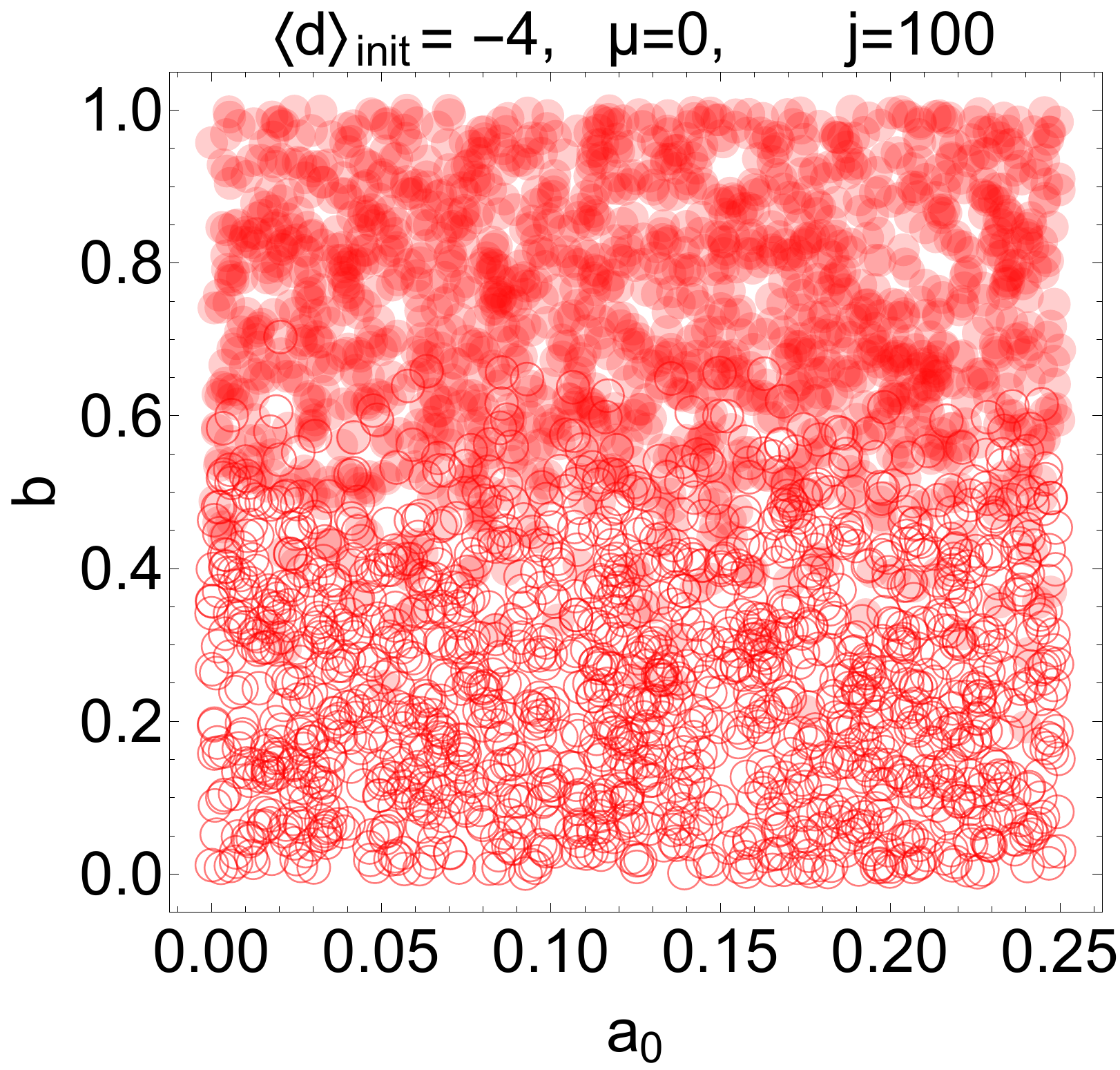}
\includegraphics[scale=0.32,trim=0  0 0 0]{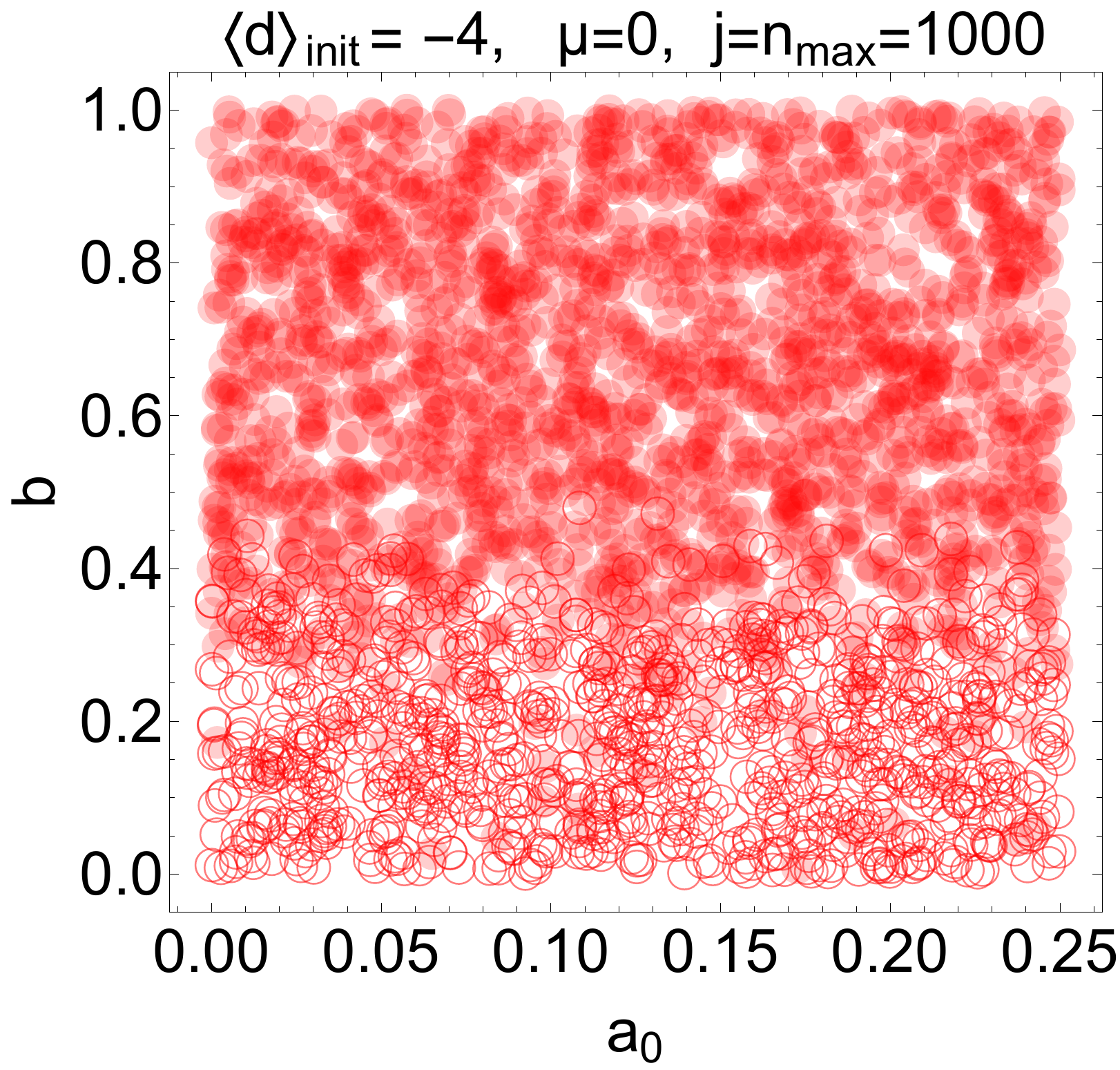}
\\
\includegraphics[scale=0.32,trim=0  0 0 0]{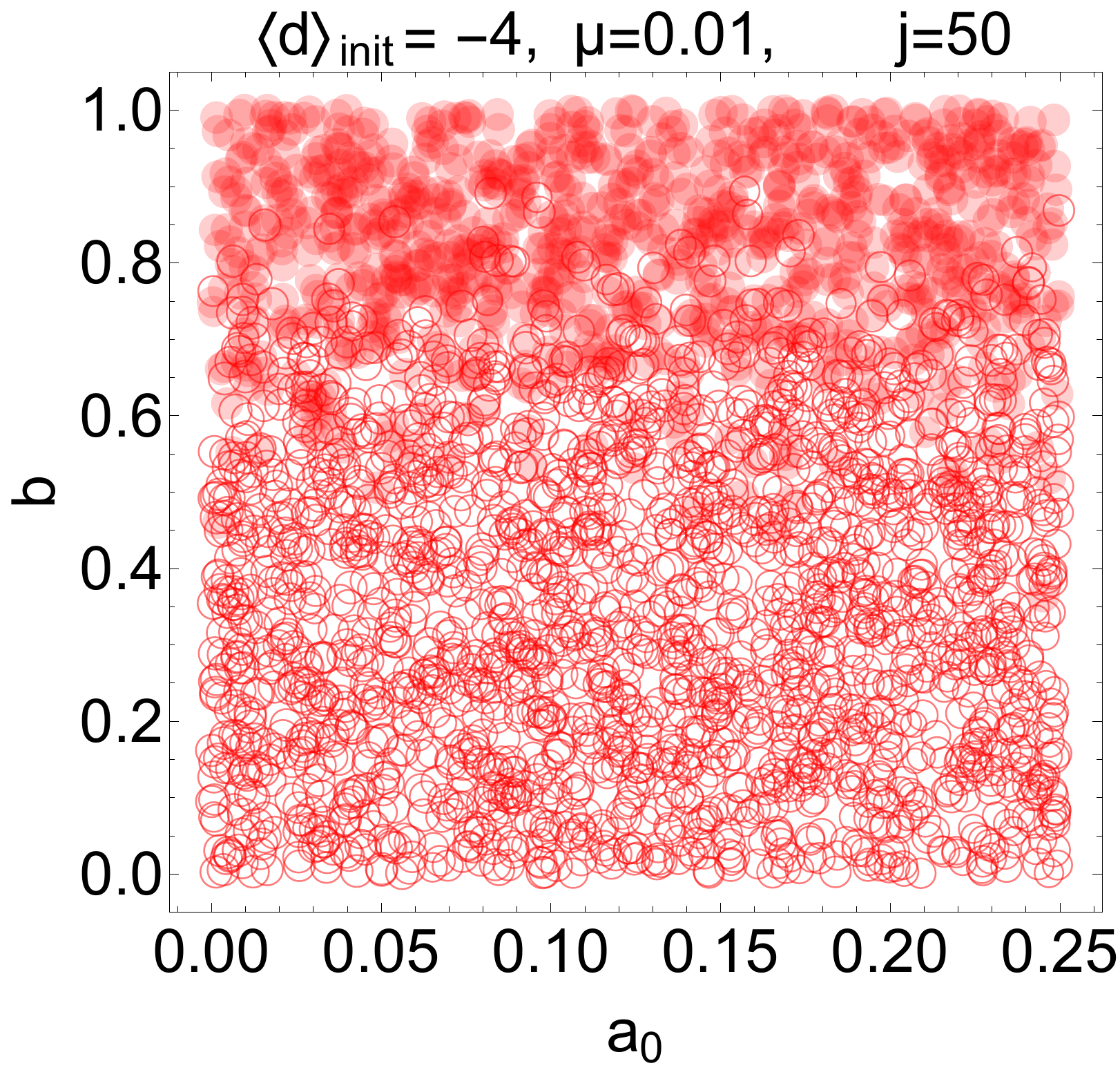}
\includegraphics[scale=0.32,trim=0  0 0 0]{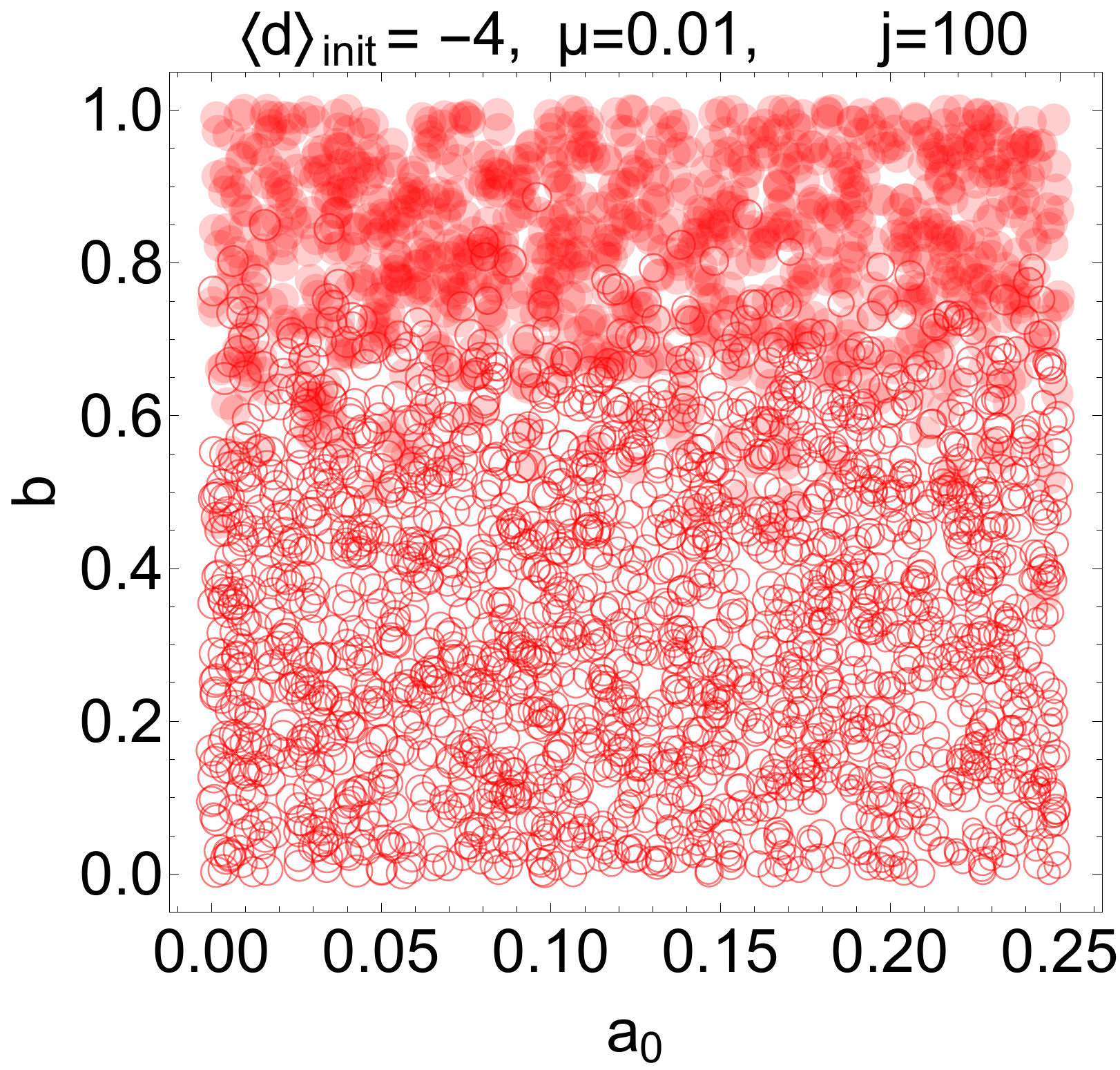}
\includegraphics[scale=0.32,trim=0  0 0 0]{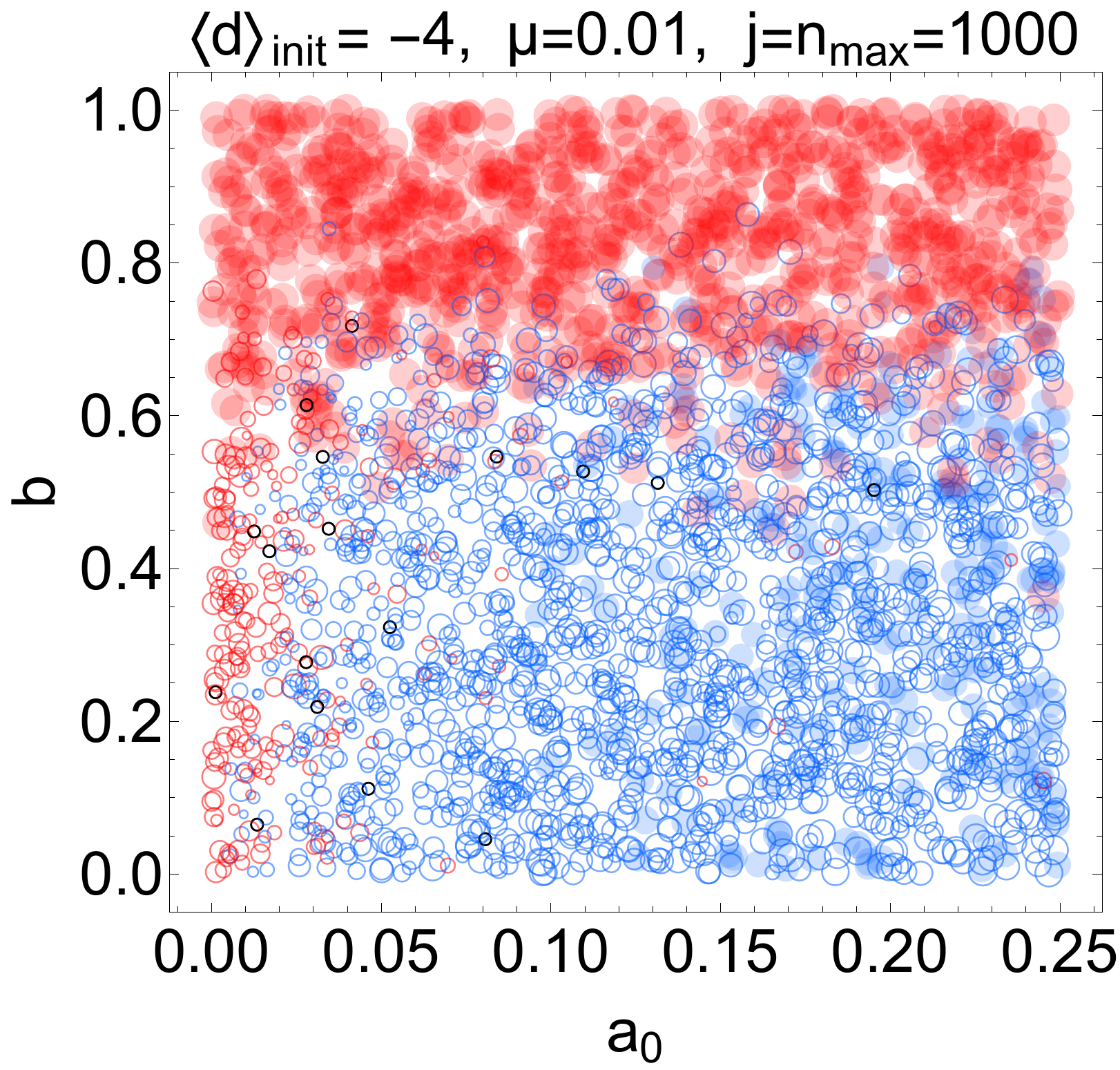}
\\
\includegraphics[scale=0.32,trim=0  0 0 0]{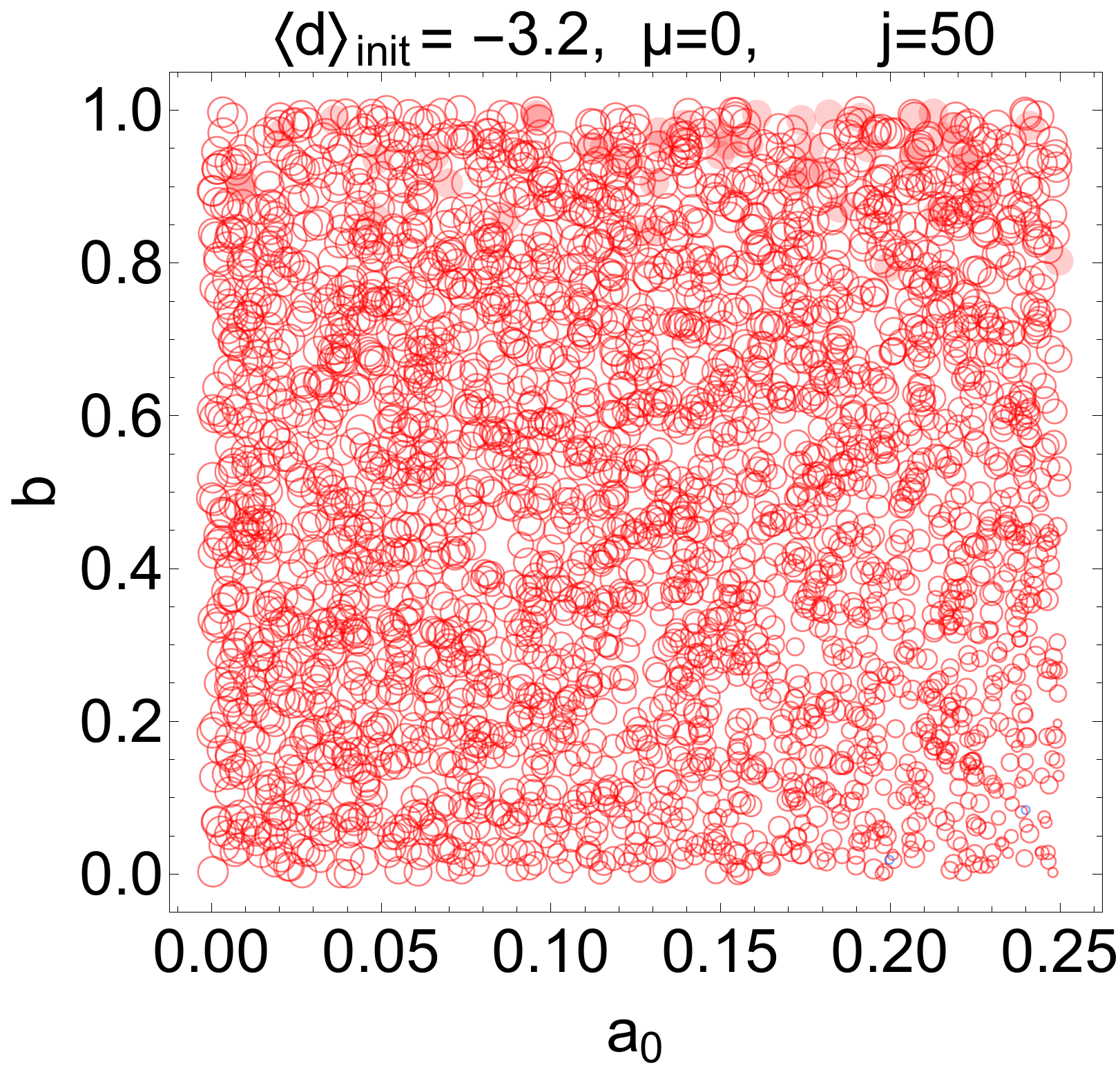}
\includegraphics[scale=0.32,trim=0  0 0 0]{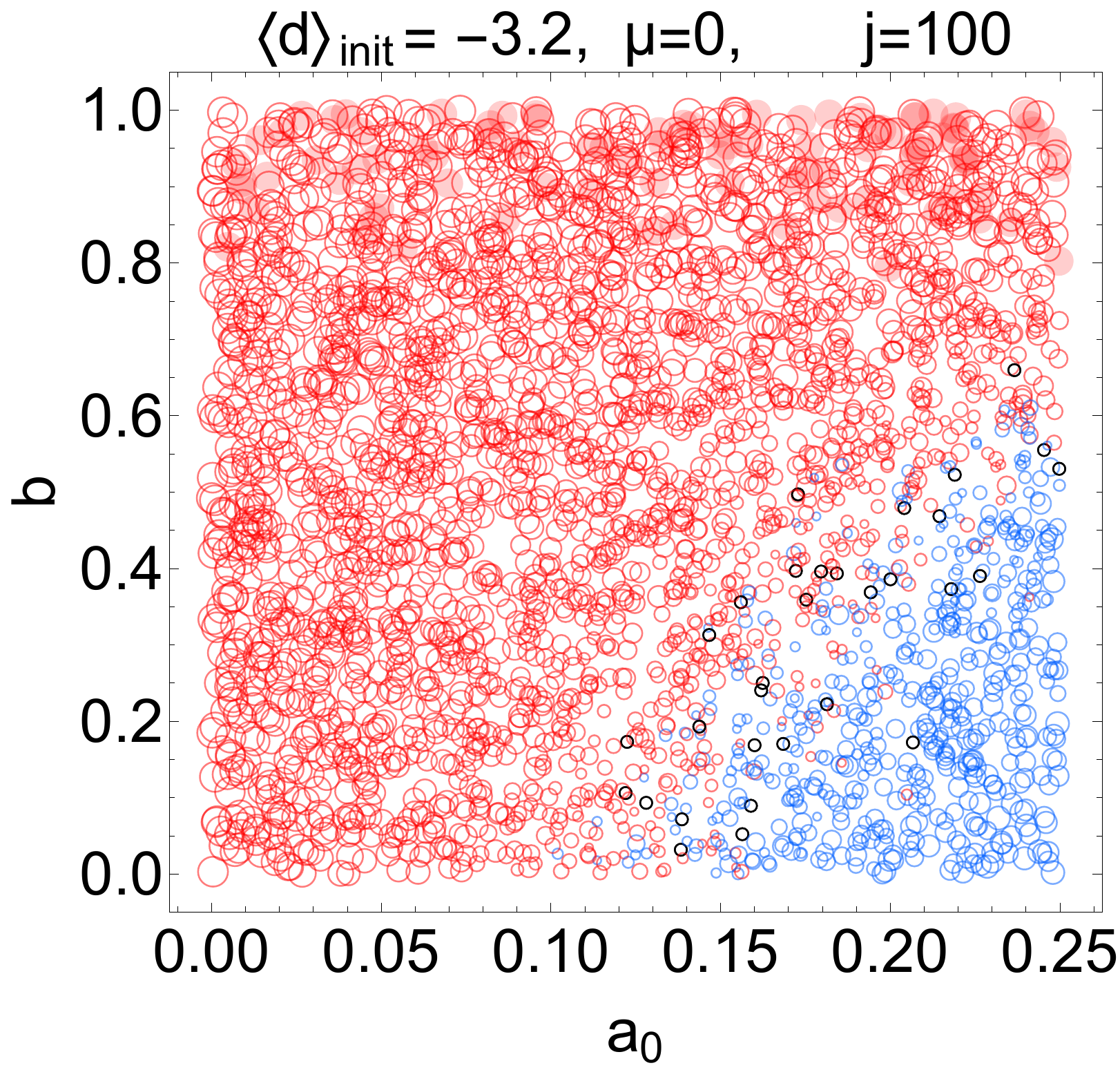}
\includegraphics[scale=0.32,trim=0  0 0 0]{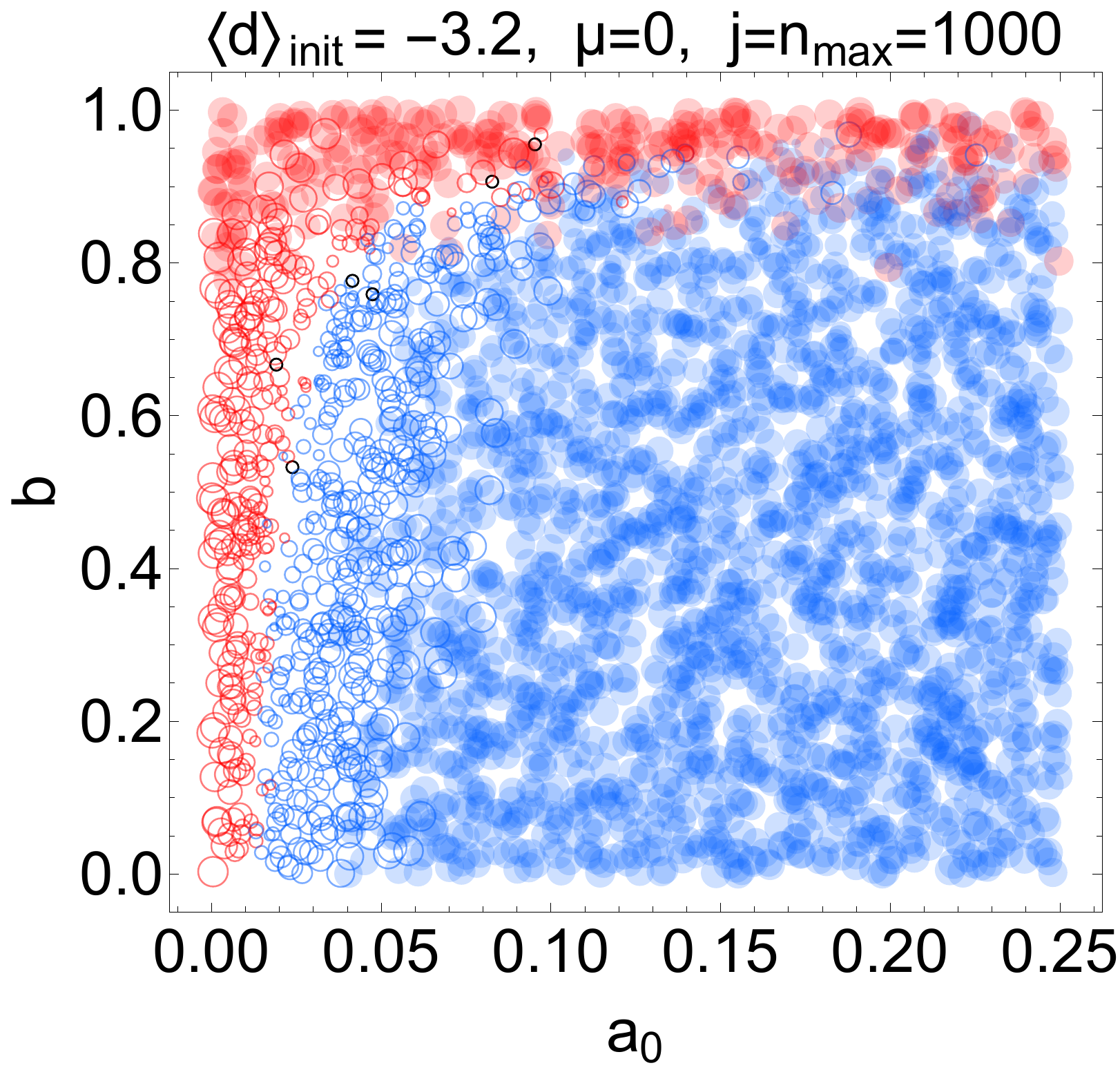}
\\
\includegraphics[scale=0.32,trim=0cm 0 0 0]{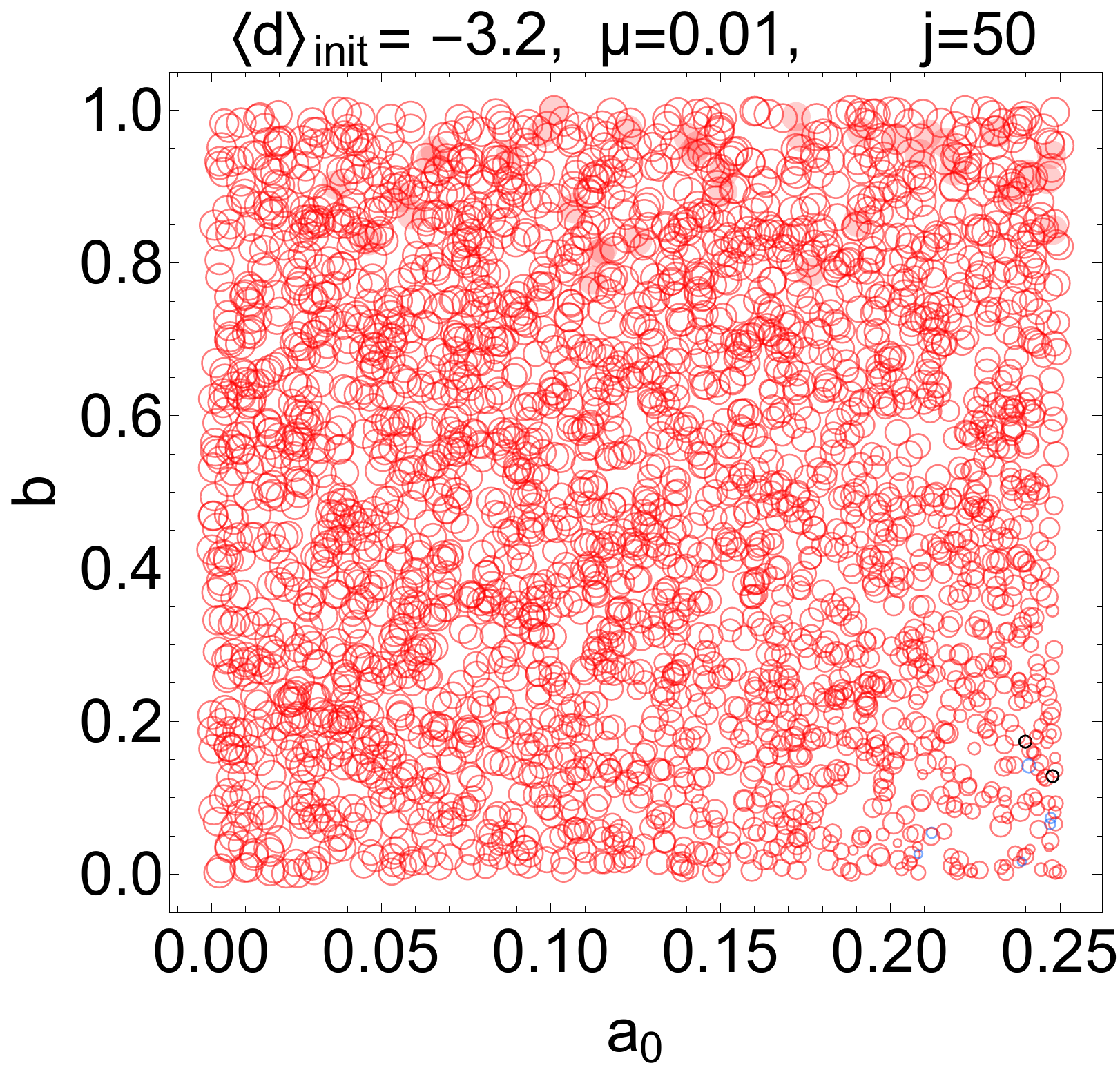}
\includegraphics[scale=0.32,trim=0cm 0 0 0]{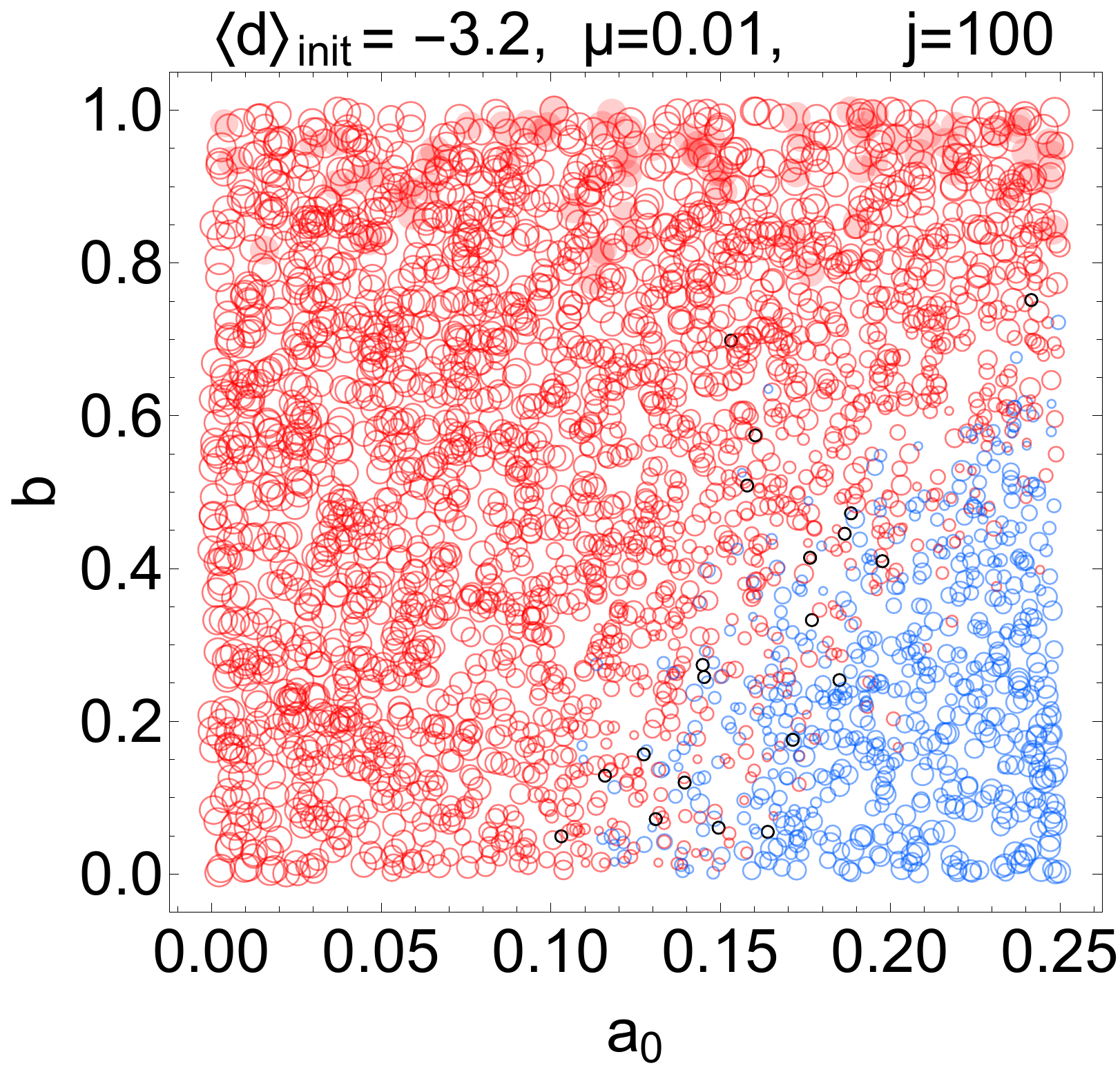}
\includegraphics[scale=0.32,trim=0.cm 0 0 0]{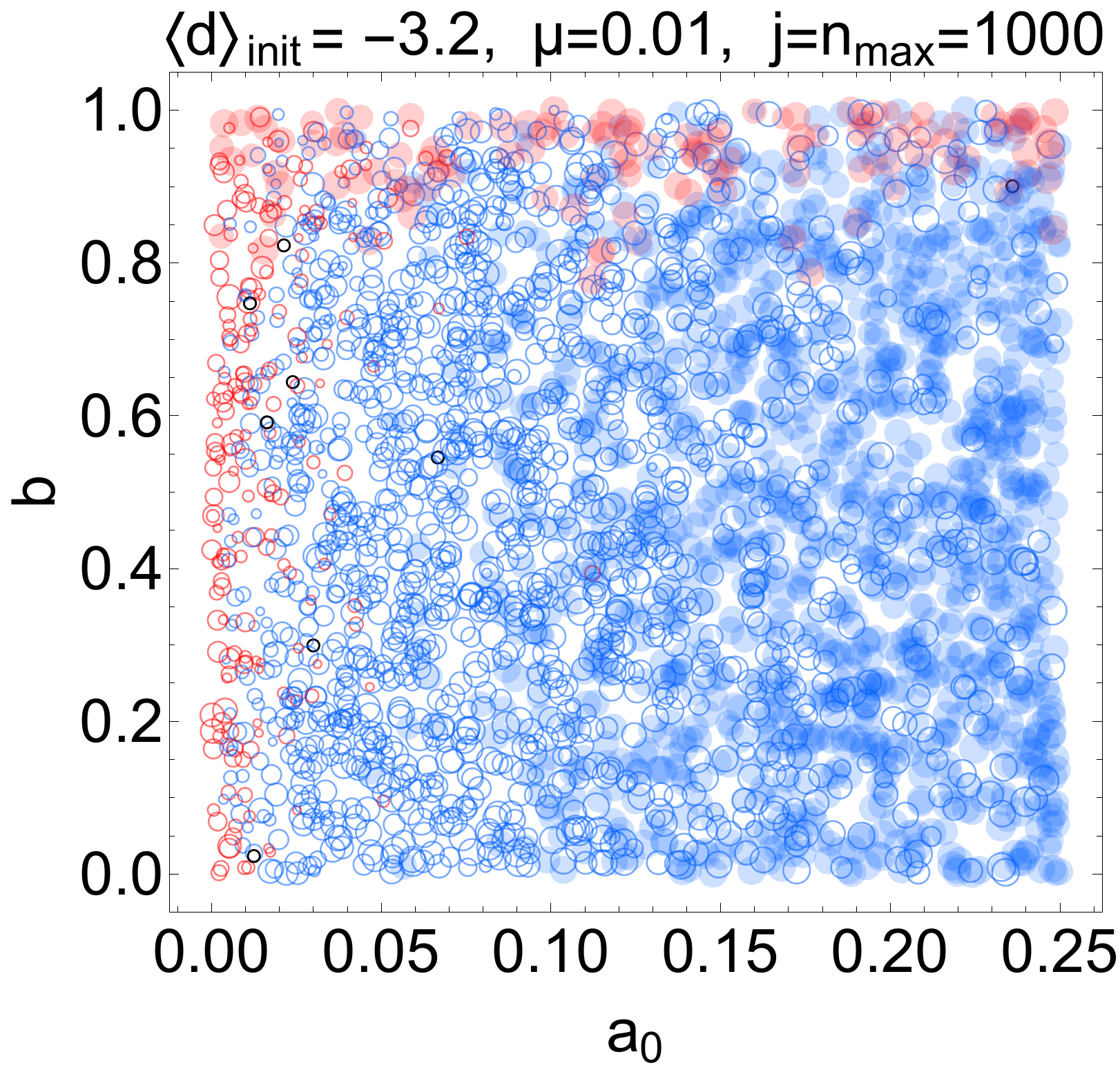}
    \caption{Top row: $\mu=0$ and $\langle d \rangle_{\text{init}}=-4$. Second row: $\mu=0.01$ and $\langle d \rangle_{\text{init}}=-4$. Third row: $\mu=0$, $\langle d \rangle_{\text{init}}=-3.2$ (18 experts with $d = -4$, 2 with $d = 4$).
    Last row: $\mu=0.01$, $\langle d \rangle_{\text{init}}=-3.2$ (18 experts with $d = -4$, 2 with $d = 4$). See text for discussion.
    } \label{f.phase_casesII_hdep} 
\end{figure*}


\section{Robustness}
\label{s.robustness}

Having just illustrated the basic operation of the reward algorithm, we briefly investigate the effects of modifying some of the assumptions used in the sample model of expert behavior above
and confirm that these do not significantly alter the main qualitative trends observed above. 

\subsection{Mutation rate}

An important parameter is the mutation rate, $\mu$, which was fixed at $0.01$ in all simulations above. The purpose of the mutation rate was to model the rather random ways that individual experts will naturally modify their views independently of external pressure or direct evidence. If the mutation rate is set exactly to zero, experts will only update their beliefs by comparing their own cumulative reward counts with those of other experts. Thus, once all experts have obtained the same degree belief $d$, there can be no more updating. A nonzero mutation rate is necessary if the group of experts to break away from an initially absolute consensus.  Increasing $\mu$ causes beliefs to migrate more rapidly, but the cost is that random variations in belief persist even as the number of predictions increases.  We illustrate these trends in \fref{phase_casesII_hdep}. Each column shows simulations at $j = 50$, $100$ and $1000$ predictions respectively. On the first row, all experts start with extreme disbelief in $\theor$ ($d = -4$), and the mutation rate $\mu$ is fixed at $0$. All circles remain red as the number of predictions increases, confirming that a mutation rate of $0$ means the experts never migrate beyond the largest $d$. The second row is the same as the first row, but now with a nonzero mutation rate $\mu = 0.01$. For large $a_0$ and small $b$, the group migrates to belief in $\theor$, albeit only after many predictions. On the third line, we return to the case of $\mu = 0$, but now we allow two of the experts to begin with strong belief ($d = 4$) in $\theor$, so that $\langle d\rangle = -3.2$ at the outset. Now the two believing experts moving the group very decisively to the $d > 0$ state unless $b$ is very large ($b \gtrsim 1.0$) or $a_0$ is very small ($a_0 \lesssim 0.01$). Since there are no random fluctuations in $d$ here, the transition between the blue and red regions is sharply defined, and the circles tend to be solid. 
Finally, on the last row we repeat the simulation of the third row but with $\mu = 0.01$. The group moves toward the $d > 0$ region again here. Indeed, there is only a small difference from the third row. Primarily, it is that the line separating the blue and red regions is slightly less clearly defined, given the increase in random fluctuations. 

To summarize, a small but nonzero $\mu$ is sufficient to guarantee that the group will migrate toward $d > 0$, even if all experts start with $d = -4$. A decisive final consensus is likely so long as $\mu$ is no larger than a few percent.  

\subsection{Threshold bias}

Experts will almost always correctly validate the outcome of a 
prediction if they have $\hat{s}_{i,j} \ll b_0$. However, for much larger surprisals they will be strongly influenced by their biases. We originally chose $b_0 = 0.7$ because this corresponds to the surprisal in a prediction with approximately 50$\%$ odds, so any surprisal larger than roughly $0.7$ may reasonably be considered ``surprising'' for a typical expert in the absence of any other considerations. However, actual experts may (and will likely) operate with different effective surprisal thresholds. A level of surprise that is acceptable to one expert may be intolerable to another. The assumption that \emph{any} suprisal above $0.7$ may be considered large is rather conservative. A more realistic scenario is that most experts can accept surprisals somewhat larger than $0.7$. 

A probability threshold of $p \gtrsim 75\%$ corresponds to $b_0 \approx 1.4$ and a probability threshold of $p \gtrsim 90\%$ corresponds to $b_0 \approx 2.3$. We expect the blue portion of the $b$ versus $a_0$ plots to increase in size if we use these larger values of $b_0$.
We confirm this by repeating the simulations corresponding to the scenario where all experts start with $d = -4$, with each of these new surprisal thresholds, and we show the results in  \fref{phase_b0dep}.  
\begin{figure*}
\centering 
    \includegraphics[scale=0.32,trim=0  0 0 0]{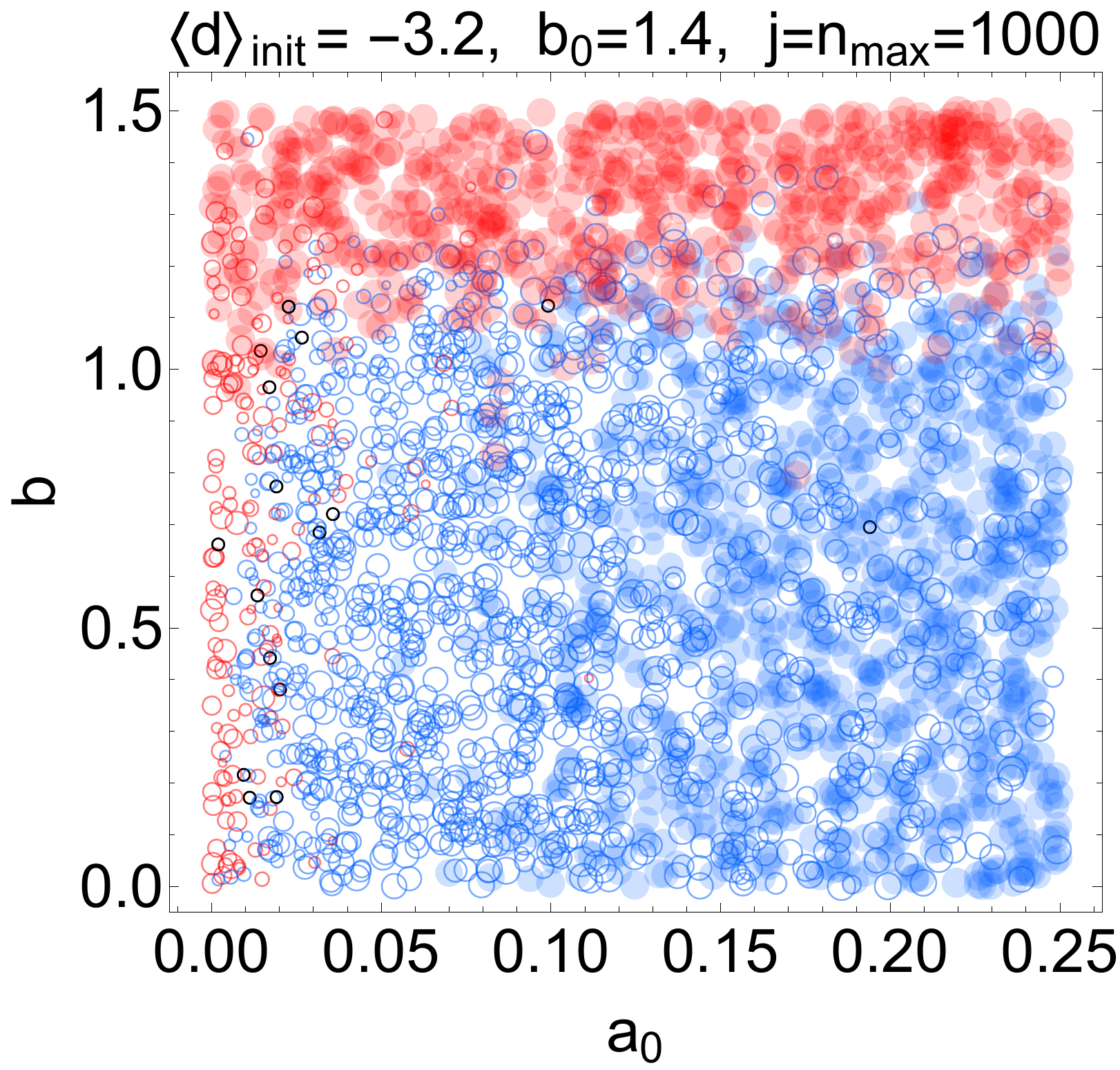}
    \includegraphics[scale=0.32,trim=0  0 0 0]{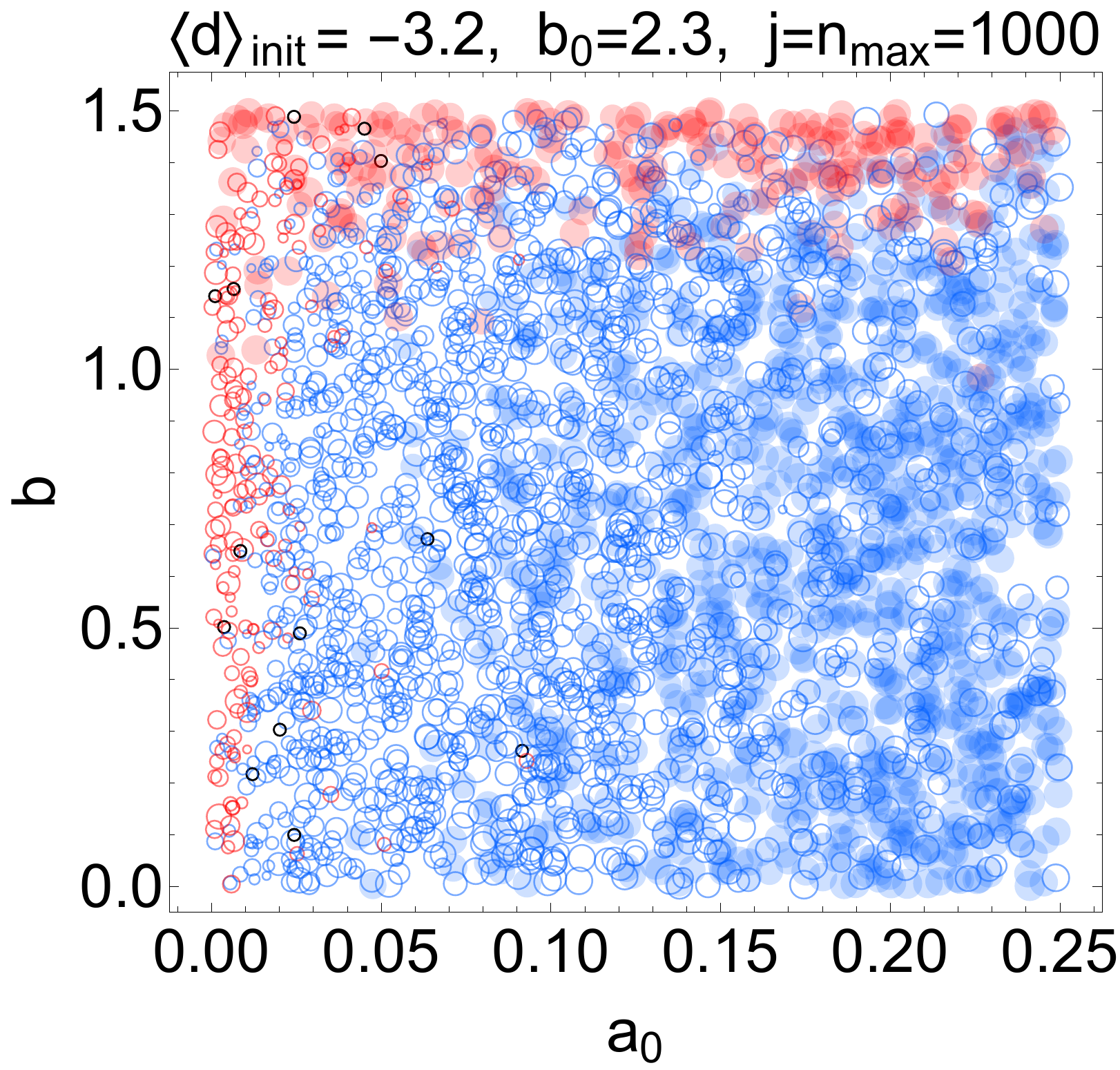}
    \caption{
    Simulations for different values of the threshold bias $b_0$, in \eref{biaseq}. In both cases, two experts start with $d=4$, and the remaining 18 with $d=-4$. The left panel shows the final state of the simulation when $b_0=1.4$. The right panel is the case $b_0=2.3$. Note the extended range in the vertical axes, respect to previous phase diagrams. As expected, increasing $b_0$ pushes the red region to larger values of $b$. These diagrams should be compared with the baseline case $b_0=0.7$, the bottom rightmost panel of \fref{phase_casesII_hdep}.}     \label{f.phase_b0dep} 
\end{figure*}
\subsection{What constitutes a ``large'' reward?}

To calculate the likelihood that experts modify their beliefs, we needed to establish a standard against which they may compare their reward counts with those of other experts. (Recall the discussion around \eref{largereward}.)
The possibilities were: i) experts compare their total reward counts with that of the group as a whole or ii) experts compare their total reward counts with that of the ``winner,'' i.e., the member of the group with the largest reward. Both options reflect influences that likely affect actual experts, and in real world groups it is likely a combination of the two, with the prevalence of each depending on the details of the specific group of experts. In our simulations, we compromised by taking a linear combination of i) and ii) and weighting each equally (see \eref{largereward}). Our last robustness test will be to examine the sensitivity to this choice of the proportion of i) and ii).  In \fref{phase_xydep}, we show the effect using the combinations $(x,y) = (3/4,1/4)$ and $(x,y) = (1/4,3/4)$ in \eref{largereward} rather than the $(x,y) = (1/2,1/2)$ that we used earlier. We have repeated the scenario in the last row of \fref{phase_casesII_hdep}, apart from the change in $(x,y)$. Namely, we start with $d = -4$ for all experts except two, who start with $d = 4$.

\begin{figure*}
\centering 
    \includegraphics[scale=0.32,trim=0  0 0 0]{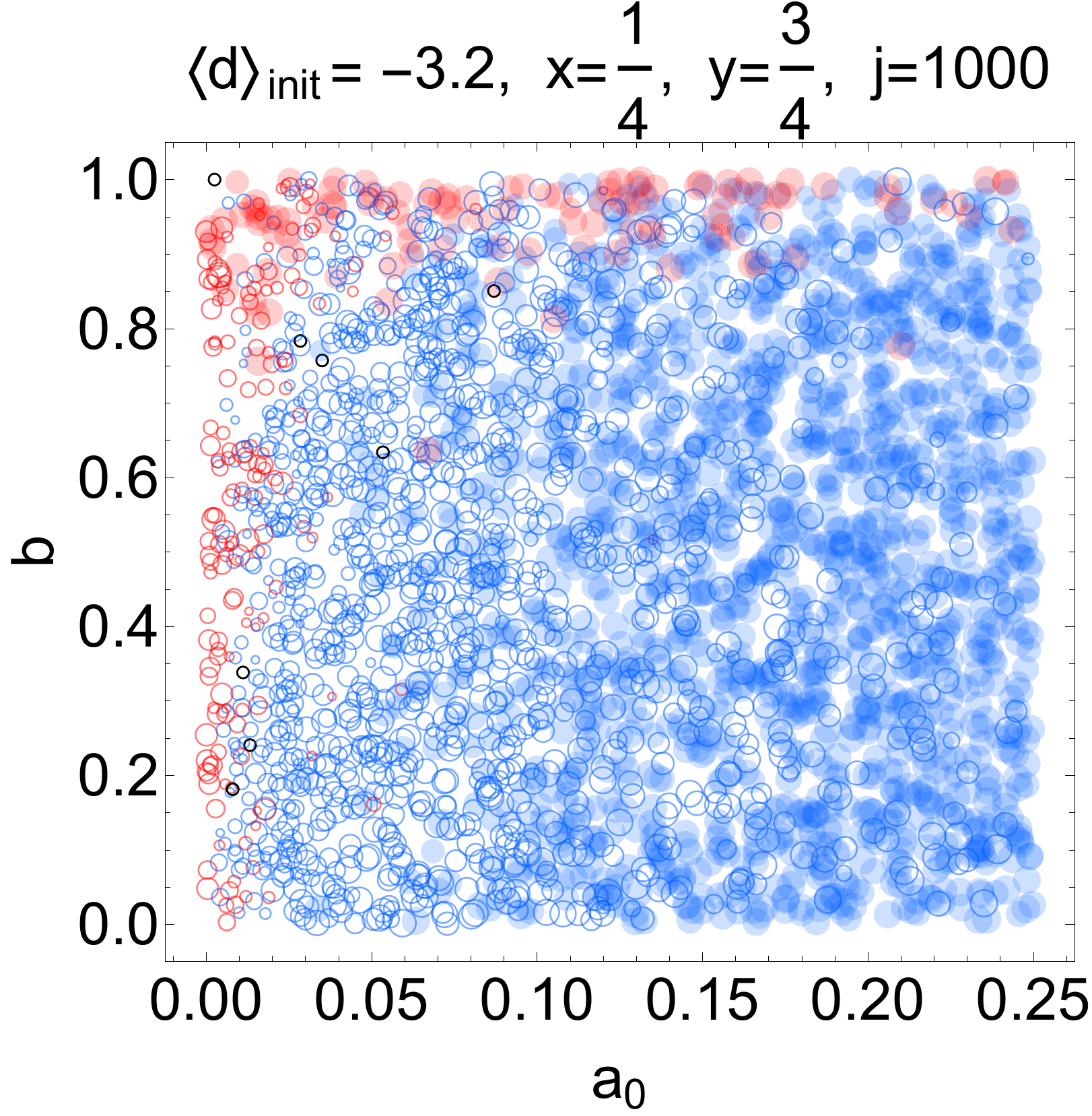}
    \includegraphics[scale=0.32,trim=0  0 0 0]{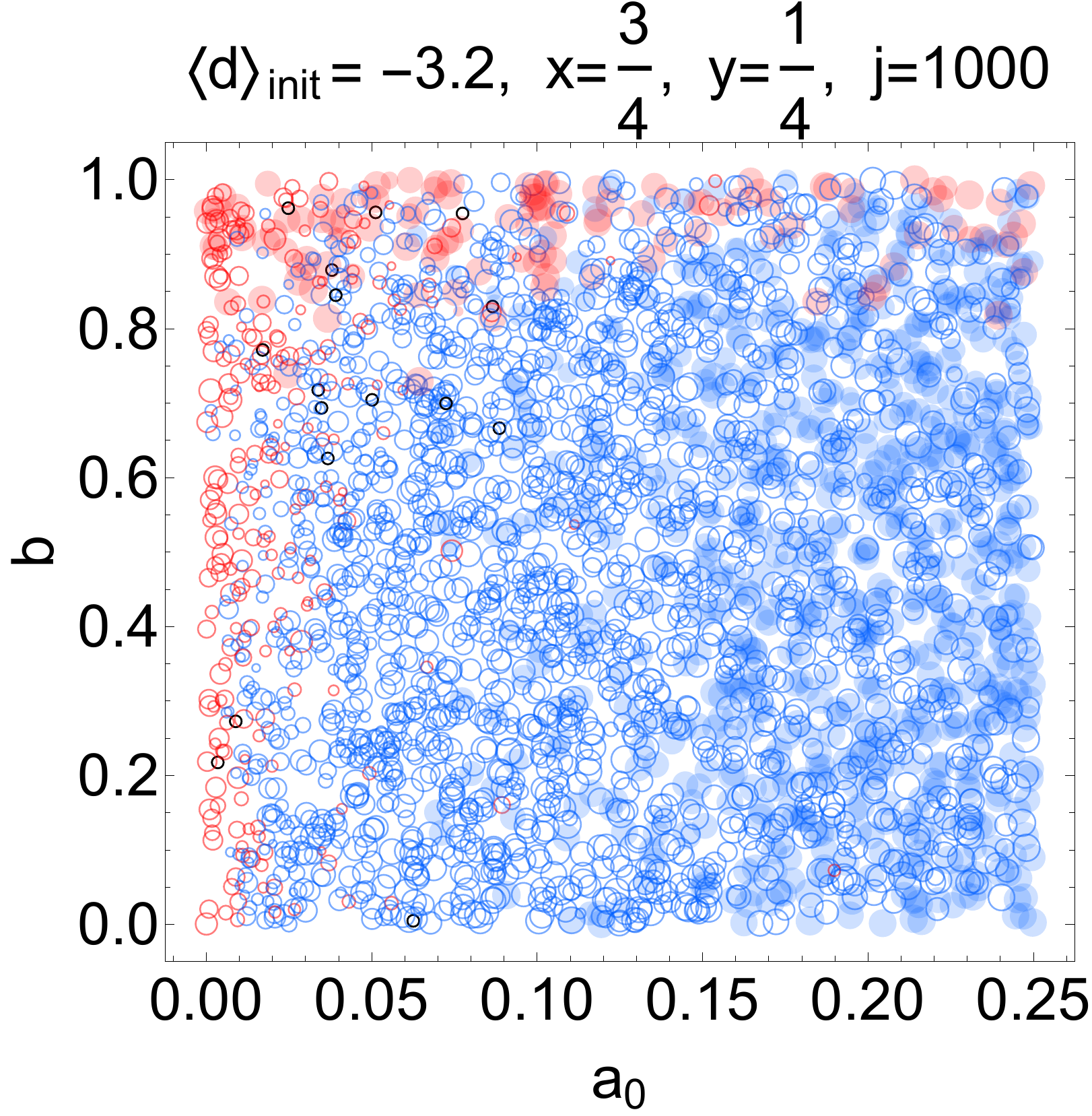}
    \caption{
    Simulations for the different values of $x$ and $y$ that enter in the definition of large rewards in \eref{largereward}. Left panel: $x=3/4$, $y=1/4$. Right panel: $x=1/4$, $y=3/4$.
    The cases shown can be compared to our baseline choice of $x=y=\frac{1}{2}$, the bottom rightmost panel in \fref{phase_casesII_hdep}.}     \label{f.phase_xydep} 
\end{figure*}

Figure~\ref{f.phase_xydep} shows that, perhaps surprisingly, there is very little effect on the qualitative behavior of the $b$ versus $a_0$ phase diagrams from modifying the exact values of $x$ and $y$

\section{Discussion and future directions}
\label{s.discussion}

The simulations above 
effectively illustrate the main principles behind the reward algorithm proposed in \eref{reward_formula}, and they show that it performs as designed under reasonable assumptions concerning the model of the expert group. If the group is collectively driven to gather large reward points (as expressed through a large $a_0$), and has a low bias (small $b$), then the group migrates toward the factually correct theory $\theor$. The important point for our purposes is that this is driven by the collective interaction between the participants and does not require an external referee to determine the true outcome or relevance of each prediction. It is in this sense that the reward algorithm is ``self-governing.'' 
To reproduce the simulations, the relevant Wolfram Mathematica documents may be found $\href{https://drive.google.com/file/d/1CE4WNy9XRZHF94WN3XkRW5kGkalddsLl/view}{here}$.

Naturally, there is still a great deal of stress-testing that still needs to be performed. One way to do this is to add complexity to the model of the interaction between experts. For example, it may be instructive to consider more complex models, \erefs{affinity}{biaseq}, for the probability distributions that determine whether experts switch their beliefs or fail to reach a consensus. Furthermore, the group of experts might be made more realistic by assigning a bias and affinity to each individual expert rather than to the group of experts as a whole. One may then investigate the effect of increasing both the magnitude of the bias and the number of individuals with a large/small bias. One of the limitations of our simulations so far is the relatively small number of experts ($20$) involved in the simulated prediction competition. 
Further insight might be gained by increasing this number. 

We intend to implement these extensions in future updates. Our long term goal for these simulations is that they become useful for guiding modifications, refinements or other updates to the algorithm in \eref{reward_formula}.






\endparano

\section{Acknowledgments} 
This work was supported by a Program for Undergraduate Research and Scholarship (PURS) grant from the Office of Research and Perry Honors College at Old Dominion University, Norfolk, Virginia, USA. We especially thank Yaohang Li, Lucia Tabacu, and Balsa Terzic for many useful discussions that helped guide this work.



% ...




 
\bibliographystyle{jasss}
\bibliography{bibliography} 


\end{document}